\documentclass[%
 reprint,
 amsmath,amssymb,
 aps,
]{revtex4-2}

\usepackage{graphicx}
\usepackage{dcolumn}
\usepackage{bm}

\usepackage{color}
\begin{document}


\preprint{APS/123-QED}
\preprint{APS/123-QED}
\title{Emergent decoherence induced  by quantum chaos in a many-body system: A Loschmidt echo observation through NMR}

\author{C. M. S\'anchez}
\author{A. K. Chattah}
\author{H. M. Pastawski}
\affiliation{Facultad de Matem\'atica, Astronom\'{i}a, F\'{i}sica y Computaci\'on - Universidad Nacional de C\'ordoba, 5000 C\'ordoba, Argentina}
\affiliation{Instituto de F\'{i}sica Enrique Gaviola (CONICET-UNC),  5000 C\'ordoba, Argentina}


\begin{abstract}
In the long quest to identify and compensate the sources of decoherence in many-body systems far from the ground state, the varied family of Loschmidt echoes (LEs)  became an invaluable tool in several experimental techniques. A LE  involves a time-reversal procedure to assess the effect of perturbations in  a quantum excitation dynamics. However, when addressing macroscopic systems one is repeatedly confronted with limitations that seem insurmountable. This led to formulate the \textit{central hypothesis of irreversibility} stating that the time-scale of decoherence, \(T_3\), is proportional to  the time-scale of the many-body interactions we reversed, \(T_2\). We test this  by implementing two experimental schemes based on Floquet Hamiltonians where the effective strength of the dipolar spin-spin coupling, i.e. \(1/T_2\),   is reduced by a variable scale factor $k$.   This extends the perturbations time scale,  \(T_\Sigma\),   in relation to  \(T_2\).  Strikingly, we observe the superposition of the normalized Loschmidt echoes for the bigger values of  $k$. This manifests the dominance of the intrinsic dynamics  over the perturbation factors, even when the Loschmidt echo is devised to reverse that intrinsic dynamics. Thus, in the limit where the reversible interactions dominate over perturbations,  the LE decays within a time-scale,  $T_3\approx T_2/R$ with $R=(0.15 \pm 0.01)$, confirming the emergence of a perturbation independent regime. These results support the central hypothesis of irreversibility.

\end{abstract}



\maketitle
\section{Introduction}

The last two decades have shown a growing interest on the understanding of how the classical limit\cite{Zu03}, thermalization\cite{BIzSZ16} and hydrodynamic behavior\cite{PaZg+Me20,Z+Yao21} emerge from quantum dynamics in closed many-body systems\cite{PoSWi06,ADLP08, GoEi16,AhPoR21,Leb21,Leb93}. This interest is driven by new quantum technologies ranging from hetero-structures to cold atoms, NV and P1 centers in diamond, Bose-Einstein condensates, and a number of others\cite{C+BlGr16,MSLu16,B+Lu17,N+Ma18,YaNa18,Z+Ca+Lu20}.  All of them have quantum excitations, ``particles" or qubits, whose interactions can be manipulated periodically to engineer new forms of synthetic quantum matter away from its ground state.  Such progress became concatenated with the demands of quantum information and computation, and a new  theoretical and experimental drive on Many-Body Quantum Chaos (MBQC). This last became a condition to match quantum mechanics and gravity in the chaotic proximity of a black hole\cite{SeSu08,MalSSt16}.  MBQC would ensure the fast scrambling of quantum information as characterized through out-of-time order (OTO) commutators that describe an exponential increase of quantum uncertainties. Such growth could be traced back to chaotic instabilities already present in single particle dynamics\cite{LaOv69,Lau87}. Since the evaluation of these commutators requires OTO correlation functions (OTOCs) that involve a time-reversal procedure, A. Kitaev\cite{Ki17} and a number of authors\cite{Zg+Pa16,Ku18,SchS17,YCZu20}  pointed their equivalence with a family of experiments known as \textit{Loschmidt echoes} (LE)\cite{LUPa98,JaPa01,GJaPaWi12}. These implement time-reversal through the sudden inversion of the Hamiltonian sign.  Calculations and experiments in different systems confirmed the scrambling phenomenon and  dubbed it ``\textit{quantum butterfly effect}", as an initially localized information rapidly spreads and mixes up under a Hamiltonian dynamics\cite{AlFIo16,RGGa17,Li+Du17,Le+Re19,GoKur20}. 

Nevertheless,  these works have not addressed some related fundamental questions: To what degree do quantum mechanical predictions, repeatedly tested on fairly small systems, remain valid when the number of involved particles increase substantially? Are hydrodynamic behavior and equilibration just an illusion due to the coarse grained measurement? Do the systems retain their memory of the initial state, i.e. the  quantum correlations that encode it?  Most physicists would give an emphatic affirmative answer. As an example, one could invoke the widely discussed black hole information paradox\cite{Mald20} which, roughly, implies that the Hawking radiation of a black hole contains some of the information it had previously swallowed.  However, our physical intuition, and even common wisdom\cite{Hor20},  hints us on the opposite view. We are more ready to admit that when a thermodynamic limit is applicable\cite{An78}, i.e. the number of particles \(N\rightarrow \infty\) and \textit{then} the ``friction'' or energy uncertainty \(\eta\rightarrow 0\), quantum dynamics could manifest a sort of phase transition as discussed by P. W. Anderson in his insight-full paper ``More is different" \cite{An72,ZgPa17} . 
Thus, Quantum Mechanics would not be at quarrel\cite{GRW86,Wein12}, but rather the flaw might be in on how it is used. The limited availability of computational resources and analytical tools could hide a possible quantum dynamical phase transition towards intrinsic decoherence/irreversibility\cite{An72,ZgPa17}. Our approach is to search for answers to those questions using specifically planned experiments. 

 Once more, nuclear magnetic resonance (NMR)  remains at hand as a well developed tool-box to test the frontiers of quantum mechanics. Indeed, G. Feher's  electron-nuclear double resonance (ENDOR) on doped silicon, yielded the puzzling evidence  \cite{Feher1959}  that led P. W. Anderson to propose the \textquotedblleft absence of spin diffusion\textquotedblright \cite{An78},  the first quantum phase transition ever recognized\cite{Sachdev2009}. More recently, NMR was used, following another hint of Anderson\cite{An54}, to identify a quantum dynamical phase transition induced by a spin environment\cite{ADLP08}. It also allowed the observation of quantum critically\cite{Qu+ZaSu06,Z+Su08} and signatures of many-body localization on spin dynamics\cite{ASuK15,WeiChCa18}. Furthermore, a combination of magnetic resonance and optical techniques applied to impurities in diamond allowed to achieve a record in their coherence time\cite{M+CiLu12}. They also allow to address the microscopic basis for spin diffusion\cite{ZCo98,BoCo+04} by observing, in real time, the emergence of the hydrodynamic behavior\cite{PaZg+Me20,Z+Yao21}. These works involved \textit{dynamical decoupling} and other forms of LEs. The first LE  was introduced by E. Hahn\cite{Hh50,BrHh84}. His \textit{spin echo} (SE) reverts the precession of individual spins, and is limited by $T_2$, the time scale of multi-spin interactions that scramble any local excitation.  Much later, the \textit{magic echoes} (ME) achieved the recovery of that SE decay\cite{RPiWa71}.  The related  \textit{multiple quantum coherences} (MQC) \cite{MuPi86} constitute an early NMR version of the mentioned OTOCs\cite{Li+Du17, NSCo20,Sa+Pa20}. There,  many-body time-reversal procedures are repeated to quantify the number of spins effectively coupled. A further step was the \textit{polarization echo}(PE), in which a \textit{local} excitation was injected and observed to diffuse away before its partial recovery\cite{ZMEr92,LUPa98}.
  Simultaneously, we have learned to quantify the time scale  $T_\Sigma$, that characterizes the experimental errors and non-controlled interactions\cite{UPaLe98,LeChPa04}.  However,  in spite of these impressive successes we face a fundamental limitation already encountered by our predecessors\cite{ErPriv96,CoPriv97}:  many-body time-reversal fails lamentably already at rather short time-scales.  Further experiments consistently showed that the reversibility time, \(T_3\), was just a few times longer than \(T_2\)\cite{LUPa98,P00,LeChPa04,MOgBo12,NSCo20}. This seemed quite discouraging as this is the time scale of multi-spin interactions that one claims to control up to a reasonable precision of a few percent, \textit{i.e.}  \(\eta \approx 1 / T_\Sigma\ll 1/T_2\).  
 Thus, reversibility time-scale \(T_3\)  seems to be unavoidably tied to  \(T_2\). We should remark that all these solid-state NMR experiments involve a spin-lattice relaxation time \(T_1\gg \mathrm{max}(T_\Sigma,T_2 )\), which ensures a fully quantum behavior. As \(N\approx 10^{23}\), our system is already infinite to all practical purposes. Thus,  the last step to take the thermodynamic limit is to sweep the system from \( T_\Sigma< T_2\) towards  \(T_\Sigma\gg T_2\).  This is what we did by implementing a \textit{multi-pulse scaled dipolar interaction} (MPSDI) sequence that  yielded a value of  \(T_3\) consistent with an emergent property\cite{Sa+Pa20}. One might still wonder whether a lucky compensation of errors might have masked the improvement of the MPSDI sequence used to yield a universal scaling curve.  This issue is what this paper brings under definitive scrutiny by developing quite robust new tools that combine  novel\cite{SaBPC17} and traditional\cite{RPiWa70} techniques to scale down the natural interactions while keeping  \(T_\Sigma\) constant. 

With the stated purpose, we introduce two experimental procedures to measure the LEs, of the magic echo type, that use a \textit{continuous wave scaled dipolar interaction} (CWSDI) either in the backward or in the forward evolution. Each of them allows to change the relative importance of the Hamiltonian interactions respect to the uncontrolled ones. This is achieved by an off-resonance irradiation that induces a Floquet effective Hamiltonian expressed as Magnus expansion. Its zeroth-order term will be our the target effective Hamiltonian with a reduced coupling constant, while the higher order terms constitute a perturbation\cite{Hb76,KuMoSa16}. More specifically, our experiments rely on a previous implementation\cite{BuSa+15,SaBPC17} that showed how off-resonance continuous irradiation generate a scaled effective Hamiltonian and, in some cases, cancel it\cite{LeChPa04}. Meanwhile, pulse imperfections and truncation terms remain roughly constant. 

This becomes equivalent to multiply the natural dipolar Hamiltonian by a scaling factor $k$ in the forward or in the backward evolution periods during the time-reversal sequence, while the elapsed time is adapted to obtain the maximal echo condition. In many aspects, this procedure resembles the ME \cite{RPiWa70} with the additional versatility of the \(k\) factor. Our present experimental findings give further support to the central hypothesis of irreversibility \cite{ZgPa17} stating that, for unbounded systems at high temperature, there is an intrinsic irreversibility  time-scale \(T_3\)  proportional to the scrambling time \(T_2\). As we will extensively discuss below, the LE decays as a logistic function, which is consistent with our hypothesis that MBQC drives a quantum dynamical phase transition towards an emergent intrinsic irreversibility.

\section{Echoes for the scaled dynamics}\label{Sec:eco}


The system is a polycrystalline sample of adamantane consisting in $N\approx 10^{23}$ nuclear spins-$1/2$, in presence of a strong magnetic field ${\bf B}_0=B_0 \hat{z}$ that results in the Larmor frequency $\omega_0=\gamma B_0$.
At room temperatures, \(\mathrm{k_B T}\gg \hbar \omega_0 \), the system is in a Boltzmann thermal state, described by the density operator $\rho (0)=I/D+\Delta\rho (0)$,
with  $\Delta\rho (0)\propto I^z\!=\!\sum_iI_i^z$ and  $D$ the dimension of the Hilbert space. As the identity does not evolve nor gives rise to a signal we will be concerned only with the deviation $\Delta\rho$. 

The secular dipolar Hamiltonian with quantization axis $z$ in the rotating frame is,

\begin{equation}\begin{split}\label{eq:Hdip}
\mathcal{H}^z_{\textrm d}
&=\sum_{i<j}d_{ij}(3I_i^z  I_j^z-{\bf I}_i\cdot {\bf I}_j) \\
&=\sum_{i<j}d_{ij}(2I_{i}^{z}I_{j}^{z}-\tfrac{1}{2}[I_{i}^{+}I_{j}^{-}
+I_{i}^{-}I_{j}^{+}]),
\end{split}\end{equation}
which in most theoretical papers is referred as XXZ. Here, the dipolar coupling strengths are $d_{ij}=(\mu_0/4\pi)\times (\gamma^2\hbar)\times(1-3\cos^2{(\vartheta_{ij})})/(2 r_{ij}^3),$ the internuclear vector $\mathrm{\bf r}_{ij}$, and  the angle between $\mathrm{\bf r}_{ij}$ and the direction of the external magnetic field is $\vartheta_{ij}$ \cite{Sl90,Er87}.  $I^{\alpha}=\sum_i I_i^{\alpha}$ (with $\alpha=x,y,z$) are the total spin operators.  These interactions define the "spreading" time scale for the dipolar dynamics, 
\begin{equation}\label{eq:T2}
T_2=\hbar/M_2  \ \ \ \mathrm{with} \ \  M_2^2=\mathrm{Tr}[H_d^z,I^y]^2/\mathrm{Tr}[I^y I^y],
\end{equation}
from the second moment of the Hamiltonian (from now on, $\hbar=1$) .
After an initial pulse the system evolves, according the details provided in the next section,  under a effective Floquet Hamiltonian of the form $k_{F}\mathcal{H}^x_{\textrm{d}}$ during a forward time $t_F$, and then under  \(-k_{B}\mathcal{H}^x_{\textrm{d}}\) during a backward time $t_B$. Thus,
\begin{equation}\label{eq:eff}
\mathcal{H}_{\textrm{F}}=k_{F}\mathcal{H}^x_{\textrm{d}} +\Sigma_ {k_{F}}\,\,\,\text{and}\,\,\,
\mathcal{H}_{\textrm{B}}=-k_{B}\mathcal{H}^x_{\textrm{d}} +\Sigma_ {k_{B}} .
\end{equation}

The factors $k_{F}$ and $-k_{B}$ modulate the natural dipolar Hamiltonian $\mathcal{H}^x_{\textrm{d}}$ in the quantization axis $x$, as presented in Refs. \cite{BuSa+15,SaBPC17}, and \(\Sigma_ {k_{F}}\) and \(\Sigma_ {k_{B}} \)account for experimental imperfections and high order truncation errors resulting from the average Hamiltonian theory or can be neglected altogether depending on the case.

The propagator of the spin system at the end of experimental time $t= t_F+ t_B$ has the form,  
\begin{equation}\label{eq:U}
\mathcal{U}_\mathrm{LE} (t)=\exp (-\mathrm{i} t_F \mathcal{H}_{\textrm{F}})\exp (-\mathrm{i} t_B\mathcal{H}_{\textrm{B}}).
\end{equation}

The NMR signal generated by the propagator of the form (\ref{eq:U}) after a final rotation pulse is, 
\begin{eqnarray} \label{eq:LE}
M(t)= && {\rm Tr}[\exp{(\mathrm{i} t_F\mathcal{H}_{F})}\exp(\mathrm{i} t_B\mathcal{H}_{B})I^z \exp({-\mathrm{i} t_B\mathcal{H}_{B}})\nonumber\\
&&\exp({-\mathrm{i} t_F\mathcal{H}_{F}})I^z]/\mathrm{Tr}[I^zI^z].
\end{eqnarray}

If one could neglect the perturbation terms \(\Sigma_ {k_{F}}\) and \(\Sigma_ {k_{B}} \) one would have a recovered signal
\begin{eqnarray}
M(t)= && \mathrm{Tr}[\exp[({\mathrm i} \mathcal{H}^x_{\textrm d}(t_F k_{F}- t_B k_{B}))]I^z \nonumber\\
&& \exp[\mathrm{i} \mathcal{H}^x_{\textrm{d}}(t_B k_{B}-t_F k_{F} )] I^z]/\mathrm{Tr}[I^zI^z]
\end{eqnarray}

The Loschmidt echo is obtained when the backward dynamics completely reverses the forward evolution, \textit{i.e.} \(\Sigma_ {k_{F}} =\Sigma_ {k_{B}}=0 \), The condition to be satisfied is
\begin{equation}
k_{F} t_F-k_{B} t_B = 0.
\label{eq:adaptado}
\end{equation}

Only in this case the signal would result in the ideal condition where \(\Sigma_ {k_{F}}\) and \(\Sigma_ {k_{B}} \) are both identically 0, $M(t)\equiv 1$.  As the previous condition is not perfectly given in experimental procedures, the  measurement of  $M(t)$, i.e. the LE intensity, quantifies the effectiveness of the reversion process and gives insights on the contributions of \(\Sigma_ {k_{F}}\) and \(\Sigma_ {k_{B}} \) to decoherence.
Then, the LE decay allows us to define a "decoherence" time scale $T^k_{3}$, as the time at which the LE is one half, $M(T^k_3) = 1/2$, that 
quantifies the time reversal imperfections in the presence of the decoherent processes \cite{Sa+Ch16,Sa+Pa20}.

\subsection{Scaling the Hamiltonian}\label{subS:escala}

To achieve the desired scaling factor $k$, we recall our  experimental development described in Refs. \cite{BuSa+15,SaBPC17}. The procedure involves the irradiation with a r.f. field in the off-resonance condition, that is: there is a difference between the Larmor and the r.f frequencies. Here, $\Omega=\gamma b_0=\omega_0-\omega$ accounts for the off-resonance, $\omega$ is the frequency of the r.f. field,  which is applied with an intensity given by  $\omega_1=\gamma B_1$ (in $rad/s$). The secular Hamiltonian in terms of the effective frequency $\omega_e=\sqrt{\omega_1^2+\Omega^2}$, in the  \textit{tilted frame} ($X,Y,Z$), is
\begin{equation}\label{eq:HH2c}
\mathcal{H}_0^Z=- \omega_e I^Z+k_{\theta} \mathcal{H}^Z_{\textrm{d}}.
\end{equation}
This ($X,Y,Z$) frame has the $Z$-axis pointing in the direction of the effective field  $\mathbf{ B}_e=b_{0} \hat{z}+B_{1} \hat{x}$ that forms an angle $\theta$ with $\mathrm{\bf B}_0$ \cite{BuSa+15,Sl90}. The angle $\theta$ determines 
the value of the scale factor $k_{\theta}$,

\begin{equation}\label{eq:ktita}
k_{\theta} = \frac{1}{2}(3\cos^2{\theta}-1).
\end{equation}

This factor can be  varied experimentally by controlling the r.f. intensity $\omega_1$ and the off-resonance $\Omega$. Note that $k_{\theta}$ can vary continuously from $1$ to $-1/2$, when $\theta$ ranges from $0$ to $\pi/2$ \cite{SaBPC17}. The special case $k_{\theta_m}=0$ is achieved for the magic angle   $\theta_m$ given by $\cos^2{\theta_m}=1/3$, leading to an average decoupling in that condition. 
\begin{figure}
    \centering
    \includegraphics[width=8cm]{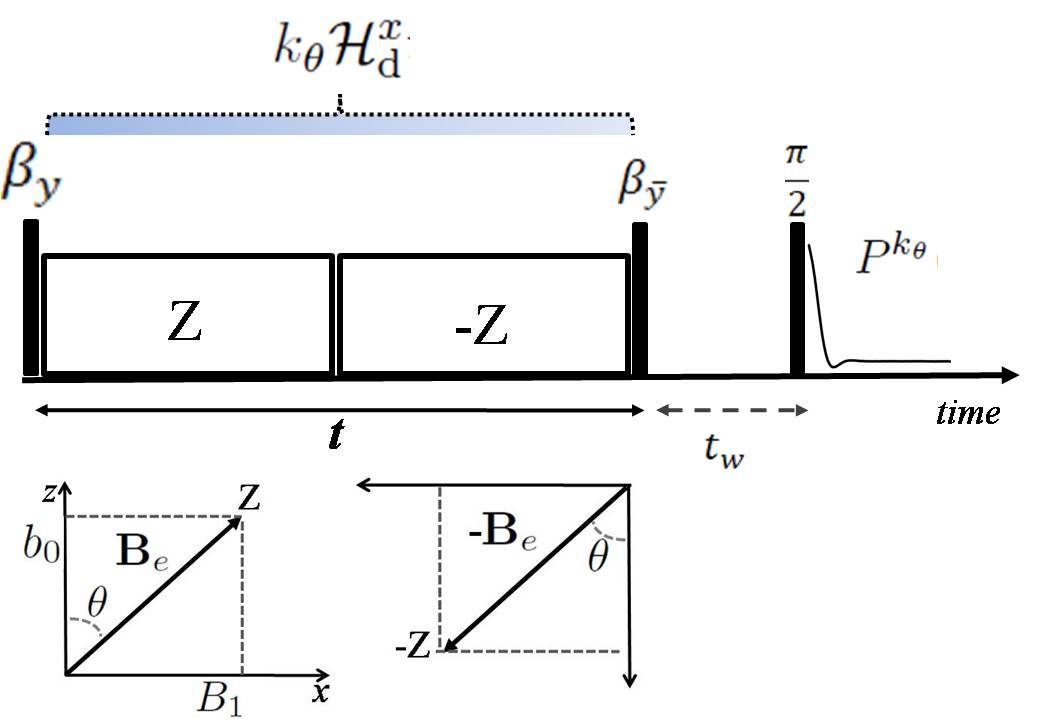}
    \caption{Experimental implementation to observe the polarization  $P^{k_{\theta}}(t)$ under the scaled dipolar Hamiltonian evolution. Schematically, the lower panel shows the directions of the magnetic fields involved in the pulse sequence (irradiation, off-resonance and effective). Each block, the irradiation time \(t_e\) is incremented in multiples of the stroboscopic Floquet time $\tau_e= 2 \pi/ \omega_e$. In a LE experiment,  \(t_e\) is chosen to compensate the natural dynamics of the other portion.}
    \label{Fig:FB}
\end{figure}
To observe the spin dynamics under the \textit{scaled} dipolar Hamiltonian, $k_{\theta} \mathcal{H}_{\textrm{d}}$ during a time $t$, we implement two successive blocks of off-resonance r.f. irradiation with effective axis $\pm Z$, surrounded by hard pulses $(\beta)_{\overline{y}}$ and $(\beta)_y$, fulfilling
$|\beta|=90^{\circ}-\theta$ (see Fig.\ref{Fig:FB}). 

Each block of duration $t/2$, multiple of the Floquet time \(\tau_e=2\pi/\omega_e\), is described by a Hamiltonian shown in Eq. (\ref{eq:HH2c}). The inversion in the phase and off-resonance of the r.f. field between the first and the second  blocks, swaps the direction $Z$ to $-Z$, leading to the reversion of the Zeeman evolution, $\pm \omega_e I^Z$.  The  effect of the $\pm \beta$  pulses, is to produce a global rotation onto the $x$-axis, yielding a propagator governed only by the dipolar term, 
\begin{equation}\label{eq:propF}
\mathcal{U}_{k_{\theta}} (t)=\exp \left(-\mathrm{i} k_{\theta}\mathcal{H}^x_{\textrm{d}}t\right),
\end{equation}
where $\mathcal{H}^x_{\textrm{d}}$ represents the dipolar Hamiltonian with the quantization axis aligned with $x$-axis of the rotating frame. The evolution of the thermal state $I^z$ under the scaled Hamiltonian, can be obtained as the signal measured by the sequence in Fig.\ref{Fig:FB}, 

\begin{equation} \label{eq:Pk}  
P^{k_{\theta}}(t)\!=\!\mathrm{Tr\!}\left[ \mathcal{U}^\dagger_{k_{\theta}} (t)I^z \mathcal{U}_{k_{\theta}} (t) I^z\right]/\mathrm{Tr}[I^zI^z]
\!.
\end{equation}

The name $k_F$ denotes the value of  $k_{\theta}$ when the irradiation angles satisfy $0\leq \theta \leq \theta_m$, leading  to a positive scaling factor in front of the dipolar Hamiltonian in the range $[0,1]$. For  irradiation angles $\theta_m \leq \theta \leq \pi/2 $, we denote  $k_B= \vert k_{\theta} \vert $ , leading to scaling factors $k_B$  in the range $[0,1/2]$ and adding a minus sign in the backward Hamiltonian. In the following the sub-index \(\theta\) will not appear, understanding that for a given scaling factor, the corresponding $\theta$ angle is set experimentally. Note also that, the extreme case $\theta=0$ with scaling factor $k_F=1$, corresponds to the natural dipolar Hamiltonian (red pulses in \textit{Scheme 1}, Fig.\ref{Fig:Scheme}) while the opposite extreme case $\theta=\pi /2$, leads a minus dipolar Hamiltonian with $k_B=1/2$ and a $Z=x$ quantization axis (on-resonance irradiation), schematized in Fig.\ref{Fig:Scheme}, blue blocks of \textit{Scheme 2} \cite{RPiWa70,RPiWa71}.
Both extreme cases do not involve off-resonance irradiation, which means less experimental error when implementing them.

\subsection{Two complementary schemes for the echoes}\label{subS:FwBw}

\begin{figure}
    \centering
    \includegraphics[width=8cm]{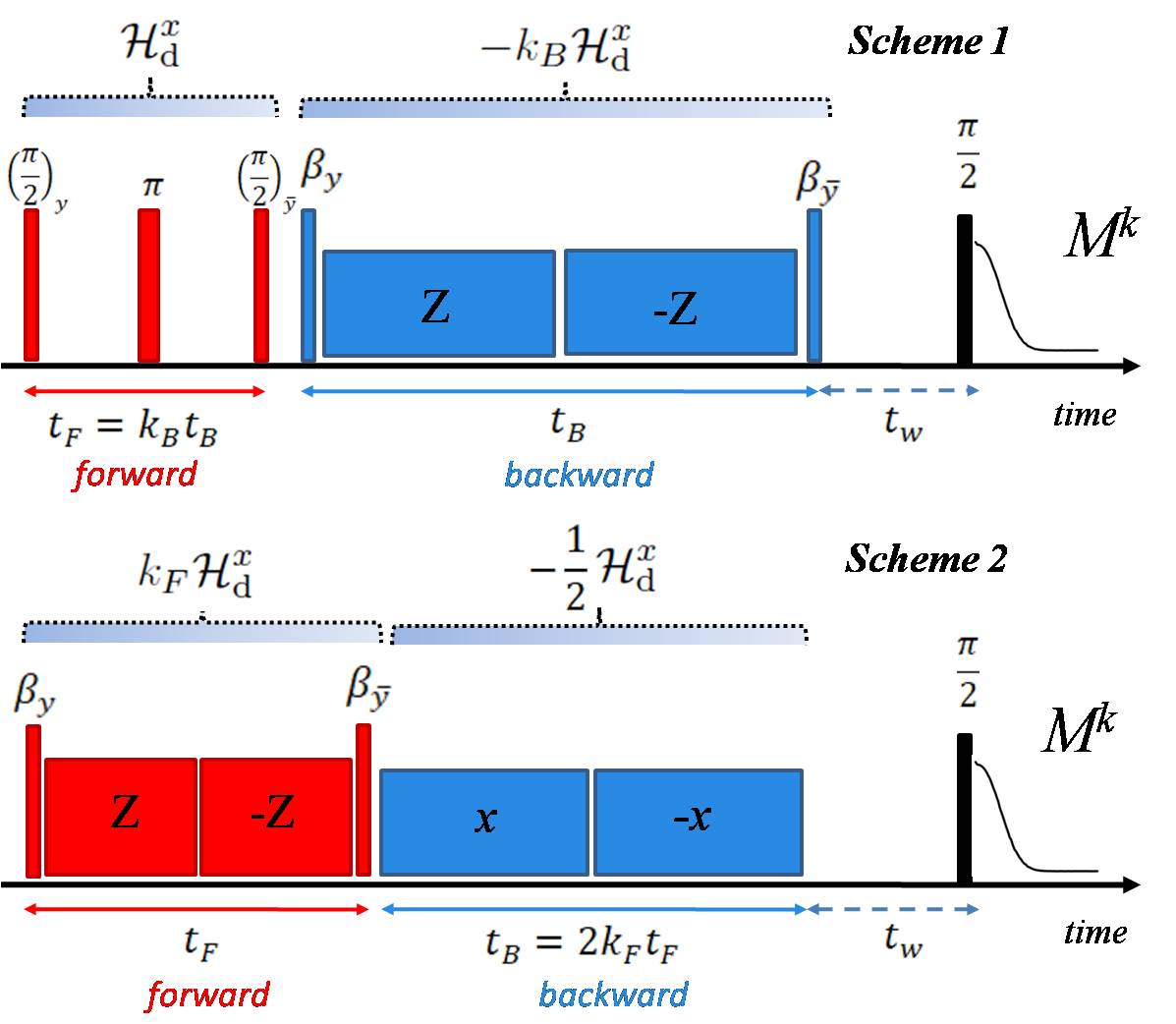}
    \caption{Two experimental r.f. pulse sequences for the observation of  the Loschmidt echo  $M^k(t_e)$. The forward evolution is red coloured while backward evolution is blue. Each block of irradiation was incremented in multiples of  the Floquet time $\tau_e= 2 \pi/ \omega_e$. \textit{\textbf{Scheme 1}}: forward dynamics is set to $k_F = 1$ and  backward is given by $-k_{B}\mathcal{H}^x_{\textrm{d}}$ 
for $k_B$ in the interval $[0,0.5]$.  \textit{\textbf{Scheme 2}} : the forward evolution is given by  $k_{F}\mathcal{H}^x_{\textrm{d}}$ for $k_F$ in the interval $[0,1]$, backward one is set with on-resonance irradiation $k_B=1/2$. The variable scale factor in \textit{Scheme 1} is $k_B$, while in \textit{Scheme 2} is $k_F$. }
    \label{Fig:Scheme}
\end{figure}

 We introduced the modulation of  the dipolar dynamics with scaling factors $k$ in ref. \cite{BuSa+15},  with the purpose to evaluate the corresponding LE decay, in a protocol dubbed  \textit{proportionally refocused Loschmidt} (PRL) \textit{echo}. That sequence used two CWSDI with identical factors $k$ in the corresponding forward and backward blocks. While the PRL echo allowed us to evaluate MQC-OTOCs \cite{SaBPC17}, the concatenation of the two CWSDI  showed uncontrolled limitations in the matching of the off-resonance frequencies that prevented from a confident comparison of the LE time scales for different \(k\).

In the present work we introduce two different schemes  for the LE implementation (see Fig.\ref{Fig:Scheme}) based on a single CWSDI dynamics. Each of them serves as a test that rules out possible errors of the LE from a MPSDI dynamics. Besides, they yield a better time resolution of the dynamics much shorter Floquet time. Both schemes also supersede the limitations previously encountered with  the PRL echo sequence. 
In the \textit{Scheme 1}  the forward evolution remains fixed as the natural dipolar one, not scaled $k_F = 1$, and thus has a negligible truncation error. The backward evolution is given by $-k_{B}\mathcal{H}^x_{\textrm{d}}$ 
for different choices of the value of $k_B$ ranging in the interval $(0,0.5]$. The forward time is adapted to meet the condition $t_F=k_B t_B$. 
In the \textit{Scheme 2}, the forward evolution is given by 
$k_{F}\mathcal{H}^x_{\textrm{d}}$
for different choices of the value of $k_F$ belonging to the interval $(0,1)$ while the backward evolution is implemented through on-resonance irradiation, which results in a fixed $k_B=1/2$ and involves truncation errors of the order \(b^2/\omega_1\).  In this case the backward time is adapted to meet the condition $t_B=2 k_F t_F$. As the scaling factor can be varied in a wider range, we can explore further possibilities than in the previous experimental conditions where we used strictly identical scaling and time for the forward and backward dynamics   \cite{BuSa+15,Sa+Pa20}.  
Both  schemes can be considered as generalized Magic Echo (ME) pulse sequences, in which time-reversal protocols adapt forward and backward times. Indeed Magic Echo (forward dynamics with a full dipolar Hamiltonian and backward dynamics with half of the dipolar interaction achieved through  on-resonance irradiation), is reproduced in the extreme value $k_B=1/2$ for \textit{Scheme} $1$. The novelty of the scaling factors, however, allows us to explore different ratios between the controlled many-body dynamics and experimental (or truncation) errors. For these last we are able to obtain a clear experimental bound by studying the LE decay for many-body dynamics when the quantization frame rotates very close the magic angle. 

In each scheme, the Hamiltonian of interest is the one associated to the  variable scale factor,  backward in \textit{Scheme 1}  and forward in \textit{Scheme 2}. Then, the discussion refers to the scale factor $k=k_B$ or $k=k_F$ in \textit{Schemes 1}  or \textit{2}, respectively. As time evolves, the scaled dipolar interaction connects the spin system entangling the quantum state with an increasingly complex character. This evolution time results, $t_e=t_B$ in \textit{Scheme 1} and $t_e=t_F$  in \textit{Scheme 2}. Then, Loschmidt echo $M^{k}(t_e)$ is analyzed in terms of the scale factor $k$.  

Note that both schemes are the same in the case $k=0$ (evolution without average dipolar dynamic), where the adapted time results zero in both cases, that is $t_F=k_B t_B=0$ for \textit{Scheme 1}, and $t_B=2 k_F t_F=0$ in \textit{Scheme 2}. Then, when $k=0$ the LE coincides with the measurement of the forward and backward dynamics, $M^{k=0}=P^{k=0}$.

\subsection{Experimental procedure}

The experiments were performed in a Bruker Avance II spectrometer operating at 300 MHz Larmor frequency,  with a $\pi/2$ pulse time set at $4$ $\mu s$. The sample temperature was controlled along the experiments at 303 K. Besides, it was not observed appreciable heating effects produced by the continuous r.f. irradiation. 

Previous settings were carried out to obtain the desired behavior in the different parts of the Loschmidt echo pulse sequence. Important care was focused on setting the r.f. irradiation fields and off-resonances to achieve accurately the same effective field frequency $\omega_e /2\pi = 79.8 \pm 0.2 $kHz for all $k$ factors. By fixing $\omega_e=\sqrt{\omega_1^2+\Omega^2}$, and $k$ through
$\frac{2k+1}{3}=\cos^2 \theta=\frac{\Omega^2}{\omega_1^2+\Omega^2}$ the experimental parameters $\omega_1$ and $\Omega$ were calculated for each $k$ as,
\begin{eqnarray}
\omega_1=\omega_e \sqrt{\frac{1-2k}{3}}\\
\Omega=\omega_e \sqrt{\frac{2(k+1)}{3}}
\end{eqnarray}. 

Corrections to these values were obtained by performing off-resonance nutations, for each $\omega_1$ and $\Omega$, to obtain the desired effective frequency. Indeed for a given $k$ the pure scaled Hamiltonian dynamics is obtained by implementing two consecutive blocks of r.f. pulses with opposite phases as depicted in Fig.(\ref{Fig:FB}) . This procedure has the objective of refocusing non desired evolution with the effective
Zeeman Hamiltonian in the tilted frame (i.e.,  $\propto \omega_e I^Z$), and attenuating the effects of the r.f. inhomogeneity. Then special attention has been put on experimental setting the $\pm \Omega$ values to obtain the desired performance. In particular, r.f. power ranged between
20 kHz and 79 kHz, while the off-resonance
took values up to 77 kHz.  Duration of $\beta$ pulses to rotate the quantization axis ranged between $0.70$ to $3.36$ $\mu s$. In all cases, the waiting time before the lecture  pulse was set to $t_w = 0.5$ ms, to allow unwanted transverse magnetization to decay. 

Forward and backward dynamics were evaluated for factors $\pm k$, including $k=0$. Each block of irradiation was incremented in multiples of $\tau_e=12.53$ $\mu s$ to be consistent with the average  Hamiltonian theory \cite{BuSa+15}, 
producing evolution times, in the range $25.06$ $ \mu s$ to $1.9$ ms. 

\textit{Schemes 1} and \textit{2} for quantifying Loschmidt Echo were successfully implemented for the different $\pm k$ values.
In \textit{Scheme 1}, the backward Hamiltonian was scaled by factors  $k=k_B=0.05,0.1,0.15,0.2,0.25,0.3,0.35,0.4,0.45$ and $0.5$, where the special case $k_B=0.5$ corresponds to the Magic Echo ~\cite{Hh50}. 
In \textit{Scheme 2}, the forward Hamiltonian was scaled by factors  $k=k_F=0.1,0.2,0.3,0.4,0.5,0.6,0.7,0.8,0.9$. In this case, scaling factors are strictly minor than $1$, $k_F<1$.    
For both schemes the experimental times were adapted in order to fulfil Eq. (\ref{eq:adaptado}).

\section{Decay rates of the scaled dynamics} \label{Sec:decays}

\begin{figure}
    \centering
    \includegraphics[width=8cm]{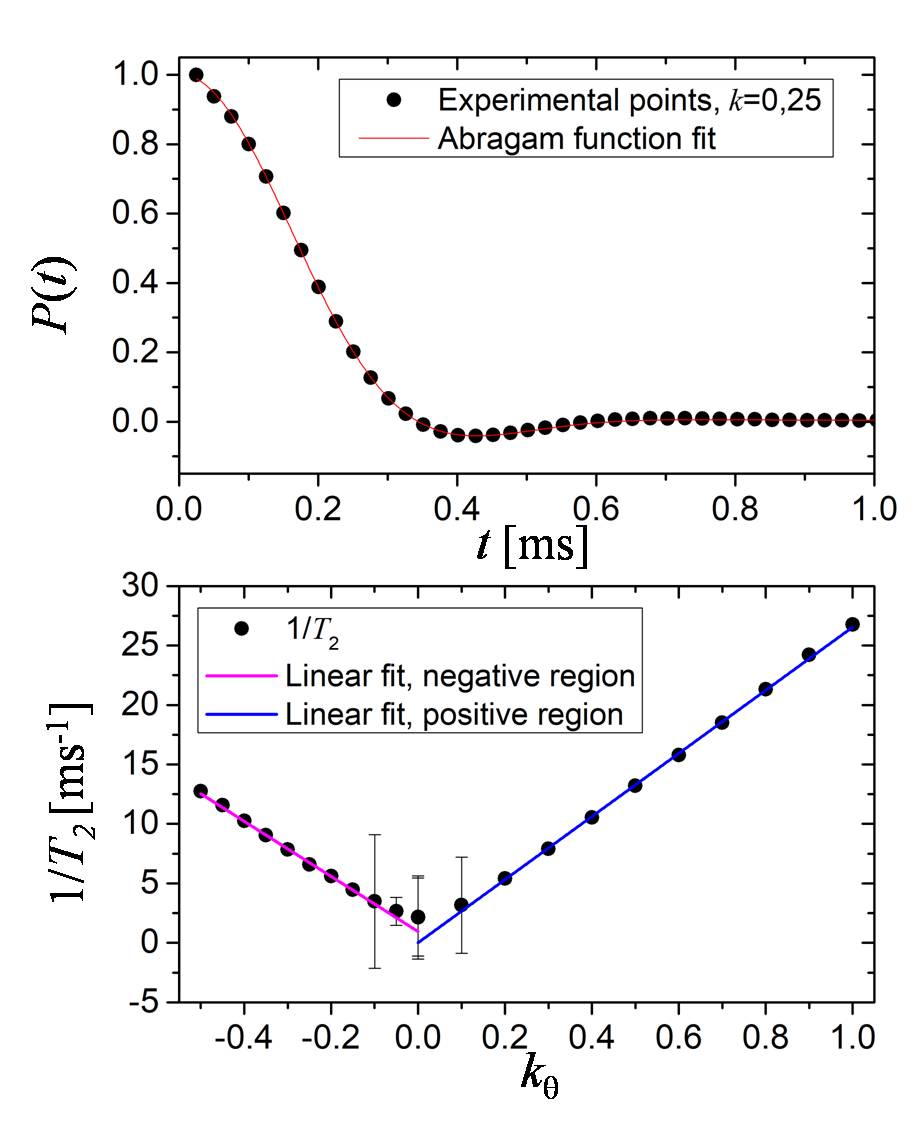}
    \caption{Upper panel:  Experimental data of $P(t)$ and  fitting to (Eq.\ref{eq:abragam}) for $k_{\theta}=-0.25$. Lower panel: $1/T_2$ obtained from the fittings to (Eq.\ref{eq:abragam}) for positive and negative values of $k_\theta$ in the range  [-0.5,1]. Linear fittings for both regions show the good performance of the pulse sequence of Fig. \ref{Fig:FB}.   }
    \label{Fig:2momento}
\end{figure}

The transverse magnetization, $I^y(t)$, under the secular dipolar    $\mathcal{H}_{\text{d}}^{z}$ (XXZ) is obtained by measuring the free induction decay (FID) after a $\pi /2$ pulse. In molecular solids as adamantane, this magnetization follows a dynamics that fits a well-known model \cite{Ab61}, the Abragam function, 
\begin{equation} \label{eq:abragam}	
P(t)=\mathrm{sinc}(wt) \times \exp[{-{(ht)^{2}}/{2}}],
\end{equation}
which captures both the decay and the damped oscillation arising from the unitary dynamics. 
The  spreading  time scale, Eq. (\ref{eq:T2}),  and the second moment of the Hamiltonian, $M_{2}=(1/T_{2})^2$, can be evaluated from the fitted parameters, $1/T_{2}=\sqrt{h^{2}+w^{2}/3}$. 

The magnetization dynamics under scaled Hamiltonian was investigated by applying the protocol of Fig~\ref{Fig:FB}. The average Hamiltonian was quenched to $ k_{\theta} \mathcal{H}_{d}^{x}$ and the total polarization as a function of the experimental time $I^z(t)$ (as expressed in Eq. (\ref{eq:Pk}))  was measured by recording the FID after $\pi /2$-pulse. 
We explored the behavior of $P^{k_{\theta}}(t)$ for various values of  $k_{\theta}$ in the range [-0.5,1). One extreme of the interval, $k_{\theta}=-0.5$, is achieved through on-resonance irradiation (blue $\pm x$ blocks in \textit{Scheme   2}, Fig. \ref{Fig:Scheme}). The limiting value $k_{\theta}=1$ is the free evolution under the secular dipolar Hamiltonian pulses of \textit{Scheme 1} in Fig. \ref{Fig:Scheme} (red on-line) , without irradiation.

Fig. \ref{Fig:2momento} displays, as example, the experimental points for $P^{k=0.25}(t)$  together with the fitting to Eq. (\ref{eq:abragam}). 
The lower panel of the same figure contains the values of $1/T^k_2$ vs. $k_{\theta}$, obtained from the fittings of $P^{k_{\theta}}(t)$ to Eq. (\ref{eq:abragam}). We observed a linear tendency of $1/T_2$ vs. $k_{\theta}$. Two different linear fittings for the positive and negative values of $k_\theta$ were performed. The slopes  were $23.0$ ms$^{-1}$ for $k_\theta > 0$ (blue on-line)  and $26.5$ ms$^{-1}$ for $k_\theta < 0$ (magenta on-line). These linear curves confirm that the pulse sequence of Fig. \ref{Fig:FB} has a very  good performance in scaling the dipolar Hamiltonian. Indeed for positive scale factors the interception of the linear curve with the ordinates is at the origin, while for negative scale factors this value is $1$ms$^{-1}$. The second moment of the scaled Hamiltonian is proportional to $k$ as expected, showing also the accuracy of the experimental value of $k$ obtained. The difference in 15 per cent in both slopes tells about experimental errors in the implementation leading to a small difference to obtain positive and negative values of a given scale factor. For negative values of $k$ the interception with \(y\) axis is not at zero. This means that there is some remaining a dipolar evolution while approaching to $k=0$ from the negative side. 

The inverse of the spreading (or spin-diffusion) times, $1/T^{k}_{2}$,  are used in following sections to analyze the behavior of decoherence vs. dynamics.

\section{Loschmidt echoes}

\begin{figure}
    \centering
    \includegraphics[width=10cm]{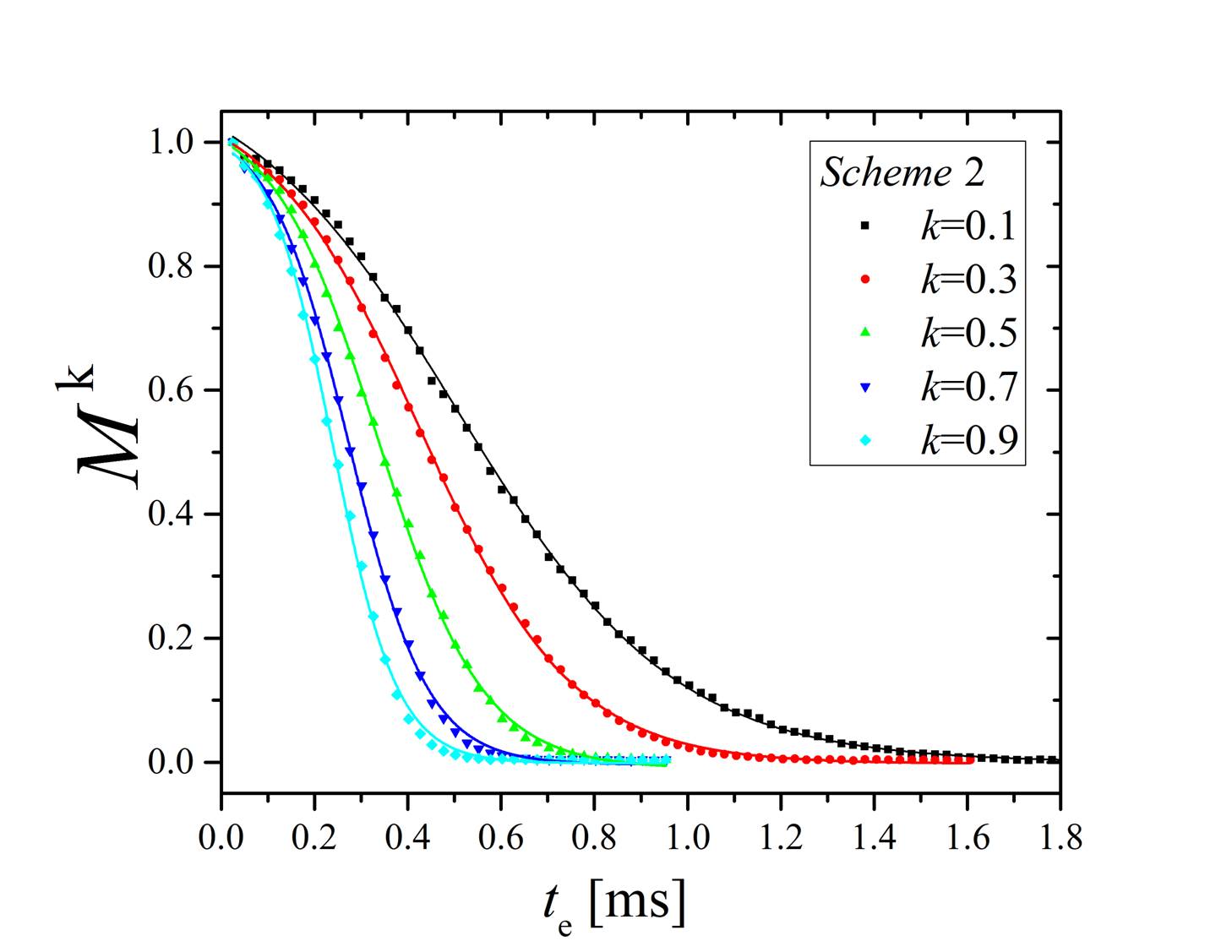}
    \caption{Loschmidt echoes for selected values of scaling factor, $M^k(t_e)$. The plot displays the experimental points normalized to the maximum value (dots), and the fittings to a logistic sigmoid function (lines). $T^k_3$ were extracted from  $M^k(T^k_3)=M^k(0)/2$. The curves show a decay monotonically ordered with $k$.}
    \label{Fig:LEsinNorm}
\end{figure}

The behavior of $M^k(t_e)$ divided by its maximum value as a function of $t_e$ can be observed in Fig. \ref{Fig:LEsinNorm}, showing a monotonic slower decay as $k$ factor diminishes. 
This behavior has been observed with another experimental setup that produced a scaled dipolar Hamiltonian \cite{Sa+Pa20}.  From these curves, the value of the decoherence times $T^k_3$ were obtained as the half height time, $M^k(T^k_3)=M^k(0)/2$.  Indeed, $M^k(t_e)$ curves follow a particular form of sigmoid, the logistic function (Fermi-function), with behavior of the form
\begin{equation} \label{eq:logistic}	
M(t_e)=C/(1+\exp[\lambda(t_e-T_3)])\ \ \mathrm{with}\ \ \  
C \approx 1.
\end{equation}
For the parameters of our system the inflection of the sigmoid is roughly the half-height time \(t_e =T_3\), and $M^k(0)=C/(1+\exp[-\lambda T_3])\approx 1$. From then on an  exponential decay occurs, with an exponent $1/\lambda \propto T_2$, that can be associated with a Lyapunov behavior that persists  as long as the signal to noise ratio is  significant. 

\begin{figure}
    \centering
      \includegraphics[width=8cm]{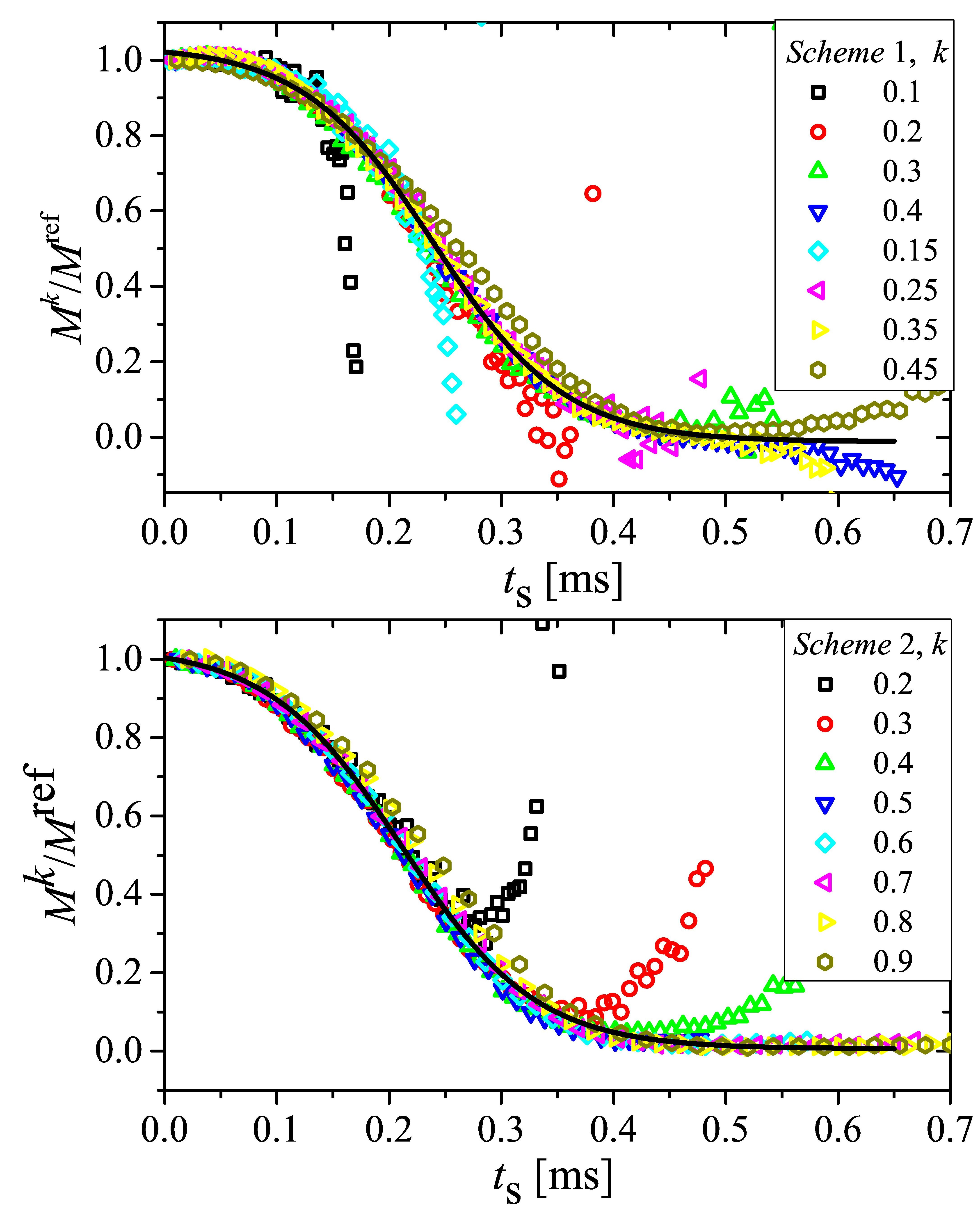}
    \caption{Normalized Loschmidt echoes as a function of the scaled time, $t_s$. Upper panel: \textit{Scheme 1}  Lower panel: \textit{Scheme 2}. Continuous lines represent the logistic sigmoid fit.}
    \label{Fig:ScaledLE}
\end{figure}

Fig \ref{Fig:ScaledLE} displays the normalized Loschmidt echoes obtained by implementing \textit{Schemes 1}  (upper panel) and \textit{2} (lower panel). Both sets of curves were plotted as a function of the scaled (or proper) time $t_s=k t_e$. The scaling factors are in the range $[0.1,0.45]$ for \textit{Scheme 1}  and in the range $[0.2,0.9]$ for \textit{Scheme 2}. The Loschmidt echo corresponding to the scale factor with lowest value (non zero) available in each scheme, was used for normalization. That is $M^{ref}=M^{0.05}$ and $M^{ref}=M^{0.1}$ for \textit{Schemes 1}  and \textit{2}, respectively. The normalization was performed by dividing for the reference at each time, $M^k(t_e)/M^{ref}(t_e)$. The selection of the lowest-non-zero $k$  for normalization has the effect to capture the basic experimental errors that underlies in each procedure, that have different implementations (off-resonance or on- resonance irradiation). Indeed the reference curves give a measure for the "perturbation" time scale $ M^{ref}(T_\Sigma)=M^{ref}(0)/2$. The normalized echoes are characterized by decays that are inherent to the coherent dynamics. This is evidenced by the superposition of the normalized Loschmidt echoes in a common curve for all values of $k$, when displaying them as a function of scaled time, as shown in Fig. \ref{Fig:ScaledLE}. Significantly, the same Figure exposes that some points of $M^k/M^{ref}$ depart from the common behavior at given times that depend on  $k$. This fact arises from the experimental errors that are accumulated differently for each $k$ and the use of different time of $M^{ref}$. Thus, the smaller the $k$ values, the longer experimental times $t_e$ are needed to observe the echo at a given scaled time $t_s$. Then, there is a departure of the common behavior that occurs monotonically with $k$, that gives a limit for the reliability of the experiments as the signal fades within the statistical noise.

\section{Decoherence vs. perturbation} 

\begin{figure}
\vspace{0.5cm}
\centering
\includegraphics[width=9cm]{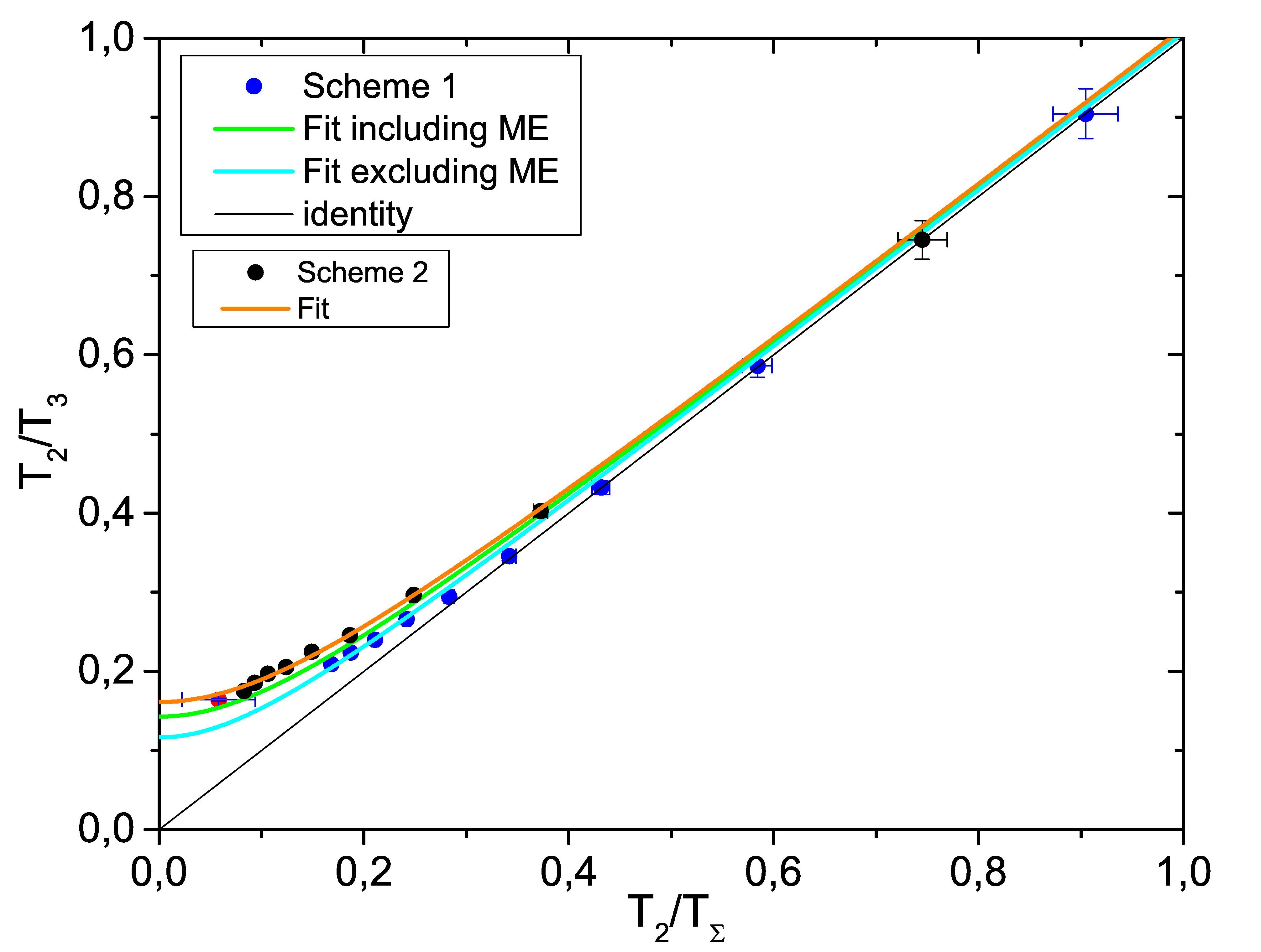}
\caption{ Decoherence rates $1/T^k_3$ in terms of the perturbation rate $1/T_\Sigma$, both divided by the scrambling rates $1/T^k_{2}$. Experimental points belong to\textit{ Schemes 1} (blue) and \textit{2} (black). A straight line with slope $1$ shows the asymptotic behavior. Fittings to  $\sqrt{A+x^2}$ were included for both schemes. 
Both schemes converge to the same ordinate at the origin, irreversibility rate \(T_2/T_3\), in spite of substantially different truncation errors in their natural Hamiltonian. 
}
\label{Fig:final}
\end{figure}

Fig \ref{Fig:final} presents the behavior of decoherence rates $1/T^k_3$ in terms of the perturbation rate $1/T_\Sigma$, through dimensionless quantities obtained dividing by the scrambling (spin-diffusion) rate $1/T^k_{2}$.
This exploration arises from the possibility to change the relative importance between coherent system dynamics and disturbance factors. The results derived from the scaling factors $k$ implemented in each  scheme were included in the figure, blue dots for \textit{Scheme 1} and black ones for \textit{Scheme 2}.  As it was mentioned before, $T^k_3$ is the inflexion points of the sigmoid fittings to $M^k(t_e)$, see Fig. (\ref{Fig:LEsinNorm}) curves, while $T^k_2$ are derived from the Abragam fitting function $P^k(t)$, displayed in Fig. (\ref{Fig:2momento}). One special mention deserves our evaluation of the perturbation rate $1/T_\Sigma$. Our lower bounds for the times $T_\Sigma$ were measured from the half-height time in the reference echoes $M^\mathrm{ref}$, \(i.e.\) the LE when the Hamiltonian vanishes. The value of $k$ selected for reference depends on the experimental implementation, as the smallest non zero scale factor  available for each scheme, that is $k=0.05$ for \textit{Scheme 1}, and $k=0.1$ for \textit{Scheme 2}. At these angles the truncation error maximizes, while still having a non-trivial dynamics.

Fig \ref{Fig:final} shows that the experimental data  $T^k_{2}/T^k_{3}$ vs. $ x=T^k_{2}/T_\Sigma$, follow a behavior of the form $\sqrt{A+x^2}$ in both schemes. Similar observation was found in Refs.~\cite{Zg+Pa16,ZgPa17}. The fitting parameter results $A=(0.020 \pm 0.001)$ for \textit{Scheme} 1, and $A=(0.026 \pm 0.001)$ for \textit{Scheme 2} . This means that
in the zone of small $k$ when  \( T_\Sigma \ll  T^k_2\), the non controlled interactions  dominate the intrinsic ones, and the experimental points fall on the line with unit slope, also shown in the figure. That is for small \(k\) one has $T_3^k\approx T_\Sigma$. In this regime $M^{k=0}(t_e)$ shows an initial quadratic (Gaussian) decay  that becomes an exponential indicative of Markovian processes \cite{UPaLe98,YaZu21}, where the best fit is obtained with an \textit{ad-hoc} function proposed by Flambaum and Izrailev\cite{FIz01}. 
In the opposite extreme, there is a region where reversible dynamics is dominant, leading to an asymptotic tendency given by $T_{2}^k/T_{3}^k=\sqrt{A}$. Then, for \textit{Scheme 1} the value $\sqrt{A}$= holds $(0.141 \pm 0.004)$ and for \textit{Scheme 2},  $(0.161 \pm 0.003)$  leading, in the limit vanishing perturbation, the universal relation $T_{2}/T_{3} \approx (0.15 \pm 0.01)$ that expresses the relation between the time-scale of irreversibility in terms of \(T_2\), the intrinsic time scale at which many-body effect manifests \cite{ZuCP07}. This value is common to both schemes, showing a universal tendency. At this point a maximum Lyapunov decay rate is reached at \(1/\lambda =1.7 T_2\). The agreement with our previous work, where the MPSDI was implemented, is quite remarkable. As the experimental implementations of a scaled dipolar Hamiltonian are based on completely different principles, the magnitude of the errors are also expected to be different. In all cases, the \(T_\Sigma\)'s were evaluated through the decay of the Loschmidt Echo for a nearly negligible dynamics  \cite{Sa+Pa20}, \textit{i.e.} \((T_2/T_\Sigma)^2 \gg 1\). 
Thus, by varying $k$ from $0$ to  $1$ (or $1/2$), the Loschmidt echo decay rate changes from the perturbation rate $1/T_\Sigma$ to the intrinsic irreversibility rate $1/T_3$, by passing through the decoherence rates. This is particularly true in \textit{Scheme 1}, since the truncation imposed by the Zeeman field is almost perfect, an decoherence occurs only during the scaled backward evolution.
Remarkably, the lower \(k\) limit coincides for both schemes. Thus, the secular dipolar interaction that yields the scrambling,  in spite of being reversed by the LE, also determines the intrinsic irreversibility rate.

\section{Connections with OTOCs and Quantum Chaos}

\textbf{Loschmidt Echoes provide a local observable.}

Some subtle points must be clarified before discussing  the physical significance of our results.  The full quantum nature of the Loschmidt Echo class of  experiments was revealed by the \textit{polarization echoes}\cite{ZMEr92,LUPa98}.
The rare nuclei of  \(^{13}C\)  nuclei were  used to inject polarization in their directly bonded \(^1H\), \textit{i.e.}  a SWAP gate\cite{ADLP06}, that also works as a detector of the polarization initiates its scrambling through the dense \(^1H\) network in a time-scale \(T_2\). 
 The normalized excitation is described by the density matrix: \(I^y=\sum_nI_n^y\), i.e. a mixed state of \textit{local} polarization at these rare protons. Since each operator in the sum constitutes a pseudo-pure state\cite{CoFH97} that evolves independently of the others, one can focus on the evolution of only \(one\) of them, say \(I^y_0\).  First, \textit{forward} in time under the many-body Hamiltonian \(\mathcal{H}\) and then  \textit{backwards} in time under  \(-(\mathcal{H}+\Sigma)\). These dynamics are concatenated in the Loschmidt Echo evolution operator \(\mathcal{U}_\mathrm{LE}(t)\)  producing a superposition of multi-spin states that conserves spin projection. Then, the total polarization \(I^y\)  at the protons connected to those rare \(^{13}C\) is \(M(t)=\mathrm{Tr}[I^y\mathcal{U}_\mathrm{LE}(t)I_0^y\mathcal{U}^\dagger_\mathrm{LE}(t)]/\mathrm{Tr}/[I_0^yI_0^y]\). After time reversal, only those many-body states that survived the perturbation and fully reconstruct the original  \(I^y_0\) can be transferred back to the corresponding \(^{13}C\).   This ensures that the actual observable is \(M(t)=\mathrm{Tr}[I_0^y\mathcal{U}_\mathrm{LE}(t)I_0^y\mathcal{U}^\dagger_\mathrm{LE}(t)]/\mathrm{Tr}[I_0^yI_0^y]\).  
 The fact that only the recovered ``individual spins'' are observed is confirmed by their interference with the polarization remaining in the  \(^{13}C\)  \cite{PLU95,LUPa98}. This results in  high-frequency oscillations (\textit{heteronuclear-coherences})  enhanced by a partial SWAP, or minimized by phase cycling. This confirms the persistence of the spin excitation quantum phase in spite of the complex many-body forward and backward dynamics\cite{PLU95,LUPa98}. 
 
 For many decades the LE of the \textit{magic echo} seemed to yield a completely different physics than the \textit{polarization echo}. On one side, neither the excitation nor the detection occur at preferred sites. On the other, its effective Hamiltonian  \textit{does not conserve} the polarization, which decays into multi-spin superpositions involving different spin projections, i.e. MQC\cite{C+CoRa05} . However,  the initial state is again a sum of independent pseudo-pure states of the form  \(I^y_0\). After time reversal, only that multi-spin component  that  reconstructs the \textit{local}  \(I^y_0\) will yield an observable signal. This is because the superposition terms that contain polarization at other sites  have phases that change wildly with time and interfere destructively during \(t_w\). This effect is confirmed by numerical simulations of many-body dynamics \cite{ADLPa10,Bend16}. Thus, a LE of the ME type, behaves much as the PE type.  Thus, all the experiments dealing  with a polarization that is not conserved:  ME,  MQC/OTOCs, and PRL-echoes, yield an observable whose locality becomes crucial for our discussion.

\textbf{On how OTOC and MQC experiments use the Loschmidt Echoes.}

In general, LE experiments considers the effect of \(\Sigma\) a \textit{persistent} ``perturbation'' or Hamiltonian imprecision, during the whole forward evolution, backward period, or both of them. 
In an ideal MQC (or the various  OTOCs),  the perturbation is just a \textit{pulse}. Let's   particularize a discussion given by Swingle\cite{Sw18}. An initial excited density matrix, say \(I^y_0=W\),  evolves for a time \(t\) under the action of a Hamiltonian, say \(\mathcal{H}\), that does not commute with it. As discussed above, \(W(t)=\exp[\mathrm{i}\mathcal{H}t]I^y_0\exp[\mathrm{-i}\mathcal{H}t]\), becomes a rather scrambled multi-spin superposition. Only  then a ``perturbation'' \(\Sigma\) acts for a very a short period \(\delta t\). One is left with  density matrix proportional to \(V^\dagger W(t)V\) with \(V=\exp[\mathrm{i}\Sigma_\theta \delta t]\). Then, a \textit{time reversed} evolution is imposed for a new period \textit{t}, producing the evolved density matrix \(\exp[-\mathrm{i}\mathcal{H}t]VW(t)V^\dagger\exp[\mathrm{i}\mathcal{H}t]\)  used to evaluate an observable, say \(W^\dagger\). In an NMR MQC experiment  \(V\) is a global rotation, such a Zeeman pulse \(V=\exp[-\mathrm{i} I^z\theta]\)\cite{Sa+Pa20}  or a field gradient pulse\cite{LoZgPa21}. Such perturbation labels with a phase \(\theta_n \) the participation of superposition state whose total spin projection differs in \(n\). Technically, one measures the total in-plane polarization D= \(I^x\). However, as pointed above, \textit{only} the component of the multi-spin state that has returned to the original site, i.e. \(I^y_0\), contributes to the observed signal, \(M_\theta (t)=\mathrm{Tr}[ W^\dagger(t)V^\dagger(0) W(t)V(0)]\). This is an OTOC that, if normalized properly, is 1 for \(t=0\) and then decays. As discussed by many authors\cite{LaOv69,Kh97, MalSSt16,Ki17,Sa+Pa20}, this decay is also related to the growth of degree of non-commutation between \(V(0)\) and \(W(t)\), as represented by the modulus square of the OTO commutator \(|[V(0),W(t)]|^2\).   In a MQC experiment, one uses the \(M_\theta(t)\) dependence on \(\theta\) to evaluate  \(|[I_0^y(t),I^y_0(0)]|^2\). This, in turn, encodes the number of spins that became entangled by the dynamics\cite{MuPi86,MuPiM87,Ki17,Sa+Pa20}. Thus, time reversal is crucial to quantify how the initial information scrambles through the Hilbert space. Indeed, some errors or Hamiltonian imprecision, as expressed by a constant part of \(\Sigma\), are experimentally unavoidable and determine the signal decay. However, this do not modify the  scrambling quantifier, i.e. the \(\theta\)-dependence for a given time. This is because one can  always normalize the data with the Loschmidt Echo,  i.e. \(M_\theta (t)/M_0(t)\). Thus, at difference from other OTOCs\cite{YoYa19}, our MQC experiments scrambling does not depend on decoherence. The drawback of the presence of \(\Sigma\) is that it imposes a limited time range\cite{ZoLu21} where the signal to noise ratio is significant . 

\textbf{Quantum chaos is the dynamical instability under a perturbed time reversal.}

One might wonder where is the Lyapunov instability, that usually characterizes chaos, manifested in our quantum mechanical experiments. A first answer needs to invoke the pioneering work of Larkin and Ovchinnikov (LO)\cite{LaOv69} and  Laughlin\cite{Lau87}. They considered semi-classical wave-packet with wavelength \(\lambda_F\) and velocity \(v_F\) that describes the conductivity of metals. The regime of ``quantum chaos''\cite{ALar96,CuPJa04} of short scattering time \(\tau_o\), is exemplified by a random array of anti-dots of size \(a\gg \sqrt{\lambda_Tv_F\tau_o}\) . There, the chaotic instability of the underlying classical system, loosely characterized by a single Lyapunov exponent \(\lambda\), manifests through a diffusion coefficient   \(D\propto v^2_F/\lambda\),  up to a logarithmic factor \(\ln[v_F\tau_o/a]\)\cite{ALar96}. This helped Kitaev to give a physical meaning to the OTO commutator as describing the exponential growth of quantum uncertainties, at least up to the Ehrenfest time \(t_E=\lambda^{-1}| \log[ a/\lambda_F]|\). While this initial exponential growth is often referred as the ``quantum butterfly effect''\cite{SchS17,Sw18},  the LE approach to quantum chaos is based on a new concept. That is, a Lyapunov instability under perturbations in the Hamiltonian during time reversal\cite{LUPa98,P00,JaPa01,ZgPa17}. This contrasts with the instability towards changes in the initial conditions which is absent in quantum mechanics, either in the simple models \cite{CaChi87} or in 1/2 spin systems\cite{ElFi15}. The most immediate effect of a weak, but persistent, Hamiltonian perturbation is a decay with a time scale \(T_\Sigma\), described by a Fermi golden rule. Thus,  increasing perturbation strengths or higher energies result in shorter \(T_\Sigma\).  However, systems with a classically chaotic correspondence showed a surprising property:  the Loschmidt Echo displayed an exponential decay with rate \(1/T_{3}=\min[{1/T_{\Sigma}, \lambda}]\) where  , \(\lambda\) is the Lyapunov exponent of the classical system. At difference from an OTO commutator growth, this exponential decay still showed up \textit{beyond} the Ehrenfest time \cite{JaPa01,JacSiBee01,JacAdBee02}. That is, as the perturbation exceeds some small critical value, \(\Sigma_c\), one  gets a classical behaviour within a fully quantum description. Indeed, a classical limit would result from the quantum description for high energies\cite{CuPJa04} where the critical value becomes negligible. In many-spin systems, the lack of a classical counterpart prevents a direct extrapolation of the discussed results. Nevertheless, the dynamics of a local excitation at high temperatures should correspond to a high energy regime and one may expect an infinitesimal critical perturbation.  In the meanwhile, it became clear that systems described by  XXZ Hamiltonians at very high temperatures, have an energy  spectrum and forward evolutions, showing signatures of quantum chaos \cite{vEvdPG94,Jy17,BIzSZ16}.  
In our case the decay of the Loschmidt echo is not a simple exponential, but Eq.\ref{eq:logistic}. Such functional form was first found experimentally\cite{RfSaPa09}, and then confirmed in a classical dynamics of a dilute gas of hard disks\cite{BeOr67} when one focuses to a local observable,  the probability that one of these disks returns to its initial neighborhood under the effect of a weak perturbation.  According to these early results, this probability decay takes the value 1/2, at a \(T_3\), that depends on the observable uncertainty and the strength of the perturbation.  The exponential decay rate  manifested at that time, was identified with the Lyapunov exponent of this classical many-body system\cite{PiMPa04}. This behavior was further observed for quantum models that describe the LE in interacting Fermions and Bose-Einstein condensates in 1D described by the Gross-Pitaevskii equation\cite{ManHe08}.  Interestingly, these last results showed that \(\lambda^{-1}\) does not depend on the perturbation strength but only on the Hamiltonian parameters. This reinforces the view that \(\lambda \propto 1/T_2\) has the role of a Lyapunov exponent. Additionally,  \(T_3\) was seen to decrease with the logarithm of perturbations strength and the system size. This last result, can be interpreted as a further indication of an emergent behaviour in the thermodynamic limit. Thus, if one accepts a rough correspondence of  such classical and quantum many-body cases with our many-spin system, we may call \(\lambda\) in Eq. (\ref{eq:logistic}) a Lyapunov exponent.

\section{Conclusions: Many-body quantum chaos leads to irreversibility.}

In the present work we have been able to evaluate the LE irreversibility time scale \(T_3\),  by  varying the ratio \(T_2/T_\Sigma\), between the time-scales of the controlled many-body Hamiltonian and that associated with the Hamiltonian imperfections. These last time scales are usually disguised, but scaling down the interactions allows its full disclosure as \(T^k_2/T^k_3\rightarrow T^k_2/T_\Sigma\) as \(k\rightarrow0\) when \((T^k_2/T_\Sigma)^2\gg 1\). In the opposite regime,  \((T^k_2/T_\Sigma)^2\ll 1\), we are approaching to the thermodynamic limit, and an asymptotic perturbation independent time scale \(T_3\) manifests the emergence of an \textit{ intrinsic irreversibility} in the many-body dynamics. The quantum chaos expresses in the LE exponential decay when irreversibility becomes intrinsic, \(T_3\simeq 6.7T_2\). Since the perturbations are  local, one needs the development of non-local superposition states before the intrinsic decoherence arises \cite{FePa15}. From then on, the LE decays exponentially with  \(\lambda^{-1}\simeq 1.7 T_2=0.25T_3\),  which cab be identified as a Lyapunov inverse rate.  This situation confirms the validity of the results observed with the MPSDI sequence\cite{Sa+Pa20}, where the nature and strength of the perturbation are different from those that appear here. This  indicates a universal characteristic of the XXZ Hamiltonians. Also the fact that \textit{Schemes 1}  and \textit{2}  coincide, definitely rules out the possibility that the truncation of the Hamiltonian could be the main source of irreversibility observed in the various forms of Loschmidt echos. This is because while, in \textit{Scheme 2} the truncation of the backward evolution is the result of the limited r.f. strength, in the \textit{Scheme 1}, the decoherent perturbations are limited only to the backward CWSDI portion. Furthermore, our present results are decisive to conclude that the scaling function is not based on an experimental artifact of the MPSDI sequences.  On the basis of this, we may recommend the evaluation of \(T_3\) by using any of the variants of MPSDI for the backward dynamics, together with a forward part in absence of irradiation.  Additionally, MPSDI could be used in biological systems without the risk of heating.

We recall that we have also implemented a Floquet Double Quantum (DQ) Hamiltonian  in the same crystal obtaining \(\lambda^{-1}\approx 0.23T_3\).  In that case the scrambling  is much faster and the number of entangled spins grows exponentially \cite{SaPaLe07,ASuK15} instead of diffusively as occurs for XXZ \cite{SaBPC17,Sa+Pa20}. The ``ballistic'' behaviour of DQ is more clear in a linear topology\cite{RfSaPa09}, showing a wave-front propagating as prescribed by the Lieb-Robinson bound\cite{PaLeU95,PUL96,Ma+Er97, CaViRa11}. The dependence of the DQ dynamics on perturbation has not yet been reported, but a multiple pulse implementation is in progress.   However, the more reversible dynamics  manifests in fact that the signal only fades away after reaching \(10^4\) entangled spins \cite{SaBPC17,ASuK15}, which largely exceeds the \(10^2\) of the dipolar case. We also may compare our present results with the LE of the polarization echo type under an XXZ Hamiltonian. There, the LE decay remains Gaussian  as long as the signal-to-noise ratio is significant. In this case, different experiments with a number of ratios between the Hamiltonian and perturbation strengths, hint an emergent \(T_3\)  of about  \( 4T_2\) \cite{P00,UPaLe98,ZgPa17}. 
As a whole, these results show that  the specific decay laws and their proportionality factors depend on the particularities of the system, like the topology of interactions network, the specific Floquet effective Hamiltonian, and of course, on the nature of the excitation. Thus, in spite of numerical support\cite{PiMPa04,ManHe08}, one should be careful not to attribute excessive universality to the observed decay laws. It is beyond our present experimental work ruling out that it could exist integrable many-body systems that manifests a perturbation independent irreversibility. Nevertheless, we accumulated two decades of experimental and theoretical antecedents on spin systems far from equilibrium in regimes that can be identified as many-body quantum chaos. 
Their general feature is that whatever a small, but global, perturbation, amplifies to allow the full reversibility. The irreversibility time scale, while proportional to the spreading time, is an intrinsic property of the experiment.

On the conceptual side, our experiments may inspire specific procedures to store and spread information. The understanding of the time scales involved could prevent strategies doomed to failure in favor of more viable alternatives. It is now clear that the original magic echoes experiments were already done in a nearly optimal regime where errors were small enough. This sets the echo decay in the perturbation independent regime, where the central hypothesis of irreversibility holds. Thus, there was no much room to improve the magic echo reversibility as its decay was determined by the intrinsic chaotic instability of the many-body system. This explains the failure of the long quest of John Waugh to improve the magic echo experiments \cite{BrHh84,CoPriv97}. On the contrary the initial flat region of the Loschmidt echo, opens the possibility to apply quantum error correction protocols while they are still effective\cite{HaPPY15}, which seems consistent with the announced plans to use over a million qubits to control the performance of a hundred qubits\cite{Hor21} . 

On a more fundamental aspect, our experiments can not represent a valid model of a quantum system that satisfy the AdS/CFT correspondence. This is because the Lyapunov exponent we found, \( \lambda \propto d/\hbar \ll\mathrm{k_B T}/\hbar\), is too small to satisfy the Maldacena bound\cite{MalSSt16} of  \( \lambda \lesssim  k_BT/\hbar\). This  quite exceptional condition is satisfied by the SYK model, especially devised with this purpose\cite{Ki17} but not by our XXZ Hamiltonians. Nevertheless, our results could provide a way out for the black-hole information paradox and to the origin of the arrow of time. Indeed, we perceive time to move in one direction only, towards the future. Yet, quantum dynamics, as other basic laws of physics, works equally well forward or backward in time. How is this apparent contradiction solved?  The more accepted view is that to account for it, we have to delve into the initial conditions after the big-bang. Indeed, Maldacena pointed that ``a black hole singularity is somewhat similar to the singularity at the beginning of the Universe, just its time reversed''\cite{Mald20}. This analysis seems reinforced by the new experiments that found that time reversal is not affected by chaos, interpreted as  \textit{the quantum butterfly noneffect}\cite{YaSi20,YaSi20sa}. There, the authors implement a sequence of operations on a qubit that could be equivalent to classical chaos. The OTOC of the quantum version, implemented in a five qubits IBM's quantum processor, showed a fine stability towards changes in the initial condition.  Indeed, there is no quarrel against these results, which are consistent with our early predictions for one-body quantum chaos\cite{JaPa01,JacSiBee01}. Our present experiments, in contrast, consider a huge number of interacting qubits, of which a few hundred become entangled. However, errors though quite small, act at every time step. The fact that we find an intrinsic irreversibility/decoherence time, adds evidence in favor of the idea that an actual chaotic many-body system far from its ground state, contains irreversibility as an emergent property of the thermodynamic limit.
The importance of such a result needs further validation that is beyond present computational possibilities.  As Feynman pointed out, the most immediate task for quantum computers is to simulate quantum systems\cite{Tra12,GANo14}. Since this is not foreseen as an easy task, it might challenge researchers that master other quantum information techniques to implement simulators of many-body LE. These should target on models, probably similar to our XXZ Hamiltonian, that  could test the emergence of intrinsic irreversibility in the thermodynamic limit.

\section{Acknowledgements}
This initial experimental work was led by Patricia R. Levstein until her passing in 2012 upon a suggestion of Richard R. Ernst in 1996. Its continuation, was stimulated by the visits to Córdoba of Alex Pines in 2002 and by R. R. Ernst in 2006.  H. M. P. acknowledges  early extensive discussions and warm hospitality at Caltech of Alexei Kitaev  in 2015, when our experimental results were still inconclusive. We acknowledge early decisive support of Fundación Antorchas, as well as grants from SeCyT-UNC, CONICET, and FoNCyT.  

\bibliographystyle{apsrev}
\bibliography{MQC-OTOC-NMR}

\begin{thebibliography}{111}
\expandafter\ifx\csname natexlab\endcsname\relax\def\natexlab#1{#1}\fi
\expandafter\ifx\csname bibnamefont\endcsname\relax
  \def\bibnamefont#1{#1}\fi
\expandafter\ifx\csname bibfnamefont\endcsname\relax
  \def\bibfnamefont#1{#1}\fi
\expandafter\ifx\csname citenamefont\endcsname\relax
  \def\citenamefont#1{#1}\fi
\expandafter\ifx\csname url\endcsname\relax
  \def\url#1{\texttt{#1}}\fi
\expandafter\ifx\csname urlprefix\endcsname\relax\def\urlprefix{URL }\fi
\providecommand{\bibinfo}[2]{#2}
\providecommand{\eprint}[2][]{\url{#2}}

\bibitem[{\citenamefont{Zurek}(2003)}]{Zu03}
\bibinfo{author}{\bibfnamefont{W.~H.} \bibnamefont{Zurek}},
  \bibinfo{journal}{Rev. Mod. Phys.} \textbf{\bibinfo{volume}{75}},
  \bibinfo{pages}{715} (\bibinfo{year}{2003}).

\bibitem[{\citenamefont{Borgonovi et~al.}(2016)\citenamefont{Borgonovi,
  Izrailev, Santos, and Zelevinsky}}]{BIzSZ16}
\bibinfo{author}{\bibfnamefont{F.}~\bibnamefont{Borgonovi}},
  \bibinfo{author}{\bibfnamefont{F.~M.} \bibnamefont{Izrailev}},
  \bibinfo{author}{\bibfnamefont{L.~F.} \bibnamefont{Santos}},
  \bibnamefont{and} \bibinfo{author}{\bibfnamefont{V.~G.}
  \bibnamefont{Zelevinsky}}, \bibinfo{journal}{Phys. Rep.}
  \textbf{\bibinfo{volume}{626}}, \bibinfo{pages}{1 } (\bibinfo{year}{2016}).

\bibitem[{\citenamefont{Pagliero et~al.}(2020)\citenamefont{Pagliero, Zangara,
  Henshaw, Ajoy, Acosta, Reimer, Pines, and Meriles}}]{PaZg+Me20}
\bibinfo{author}{\bibfnamefont{D.}~\bibnamefont{Pagliero}},
  \bibinfo{author}{\bibfnamefont{P.~R.} \bibnamefont{Zangara}},
  \bibinfo{author}{\bibfnamefont{J.}~\bibnamefont{Henshaw}},
  \bibinfo{author}{\bibfnamefont{A.}~\bibnamefont{Ajoy}},
  \bibinfo{author}{\bibfnamefont{R.~H.} \bibnamefont{Acosta}},
  \bibinfo{author}{\bibfnamefont{J.~A.} \bibnamefont{Reimer}},
  \bibinfo{author}{\bibfnamefont{A.}~\bibnamefont{Pines}}, \bibnamefont{and}
  \bibinfo{author}{\bibfnamefont{C.~A.} \bibnamefont{Meriles}},
  \bibinfo{journal}{Science Advances} \textbf{\bibinfo{volume}{6}},
  \bibinfo{pages}{eaaz6986} (\bibinfo{year}{2020}).

\bibitem[{\citenamefont{Zu et~al.}(2021)\citenamefont{Zu, Machado, Ye, Choi,
  Kobrin, Mittiga, Hsieh, Bhattacharyya, Markham, Twitchen et~al.}}]{Z+Yao21}
\bibinfo{author}{\bibfnamefont{C.}~\bibnamefont{Zu}},
  \bibinfo{author}{\bibfnamefont{F.}~\bibnamefont{Machado}},
  \bibinfo{author}{\bibfnamefont{B.}~\bibnamefont{Ye}},
  \bibinfo{author}{\bibfnamefont{S.}~\bibnamefont{Choi}},
  \bibinfo{author}{\bibfnamefont{B.}~\bibnamefont{Kobrin}},
  \bibinfo{author}{\bibfnamefont{T.}~\bibnamefont{Mittiga}},
  \bibinfo{author}{\bibfnamefont{S.}~\bibnamefont{Hsieh}},
  \bibinfo{author}{\bibfnamefont{P.}~\bibnamefont{Bhattacharyya}},
  \bibinfo{author}{\bibfnamefont{M.}~\bibnamefont{Markham}},
  \bibinfo{author}{\bibfnamefont{D.}~\bibnamefont{Twitchen}},
  \bibnamefont{et~al.}, \bibinfo{journal}{arXiv:2104.07678}
  (\bibinfo{year}{2021}).

\bibitem[{\citenamefont{Popescu et~al.}(2006)\citenamefont{Popescu, Short, and
  Winter}}]{PoSWi06}
\bibinfo{author}{\bibfnamefont{S.}~\bibnamefont{Popescu}},
  \bibinfo{author}{\bibfnamefont{A.~J.} \bibnamefont{Short}}, \bibnamefont{and}
  \bibinfo{author}{\bibfnamefont{A.}~\bibnamefont{Winter}},
  \bibinfo{journal}{Nature Physics} \textbf{\bibinfo{volume}{2}},
  \bibinfo{pages}{754} (\bibinfo{year}{2006}).

\bibitem[{\citenamefont{{\'{A}}lvarez et~al.}(2008)\citenamefont{{\'{A}}lvarez,
  Danieli, Levstein, and Pastawski}}]{ADLP08}
\bibinfo{author}{\bibfnamefont{G.~A.} \bibnamefont{{\'{A}}lvarez}},
  \bibinfo{author}{\bibfnamefont{E.~P.} \bibnamefont{Danieli}},
  \bibinfo{author}{\bibfnamefont{P.~R.} \bibnamefont{Levstein}},
  \bibnamefont{and} \bibinfo{author}{\bibfnamefont{H.~M.}
  \bibnamefont{Pastawski}}, \bibinfo{journal}{Phys. Rev. Lett.}
  \textbf{\bibinfo{volume}{101}}, \bibinfo{pages}{120503}
  (\bibinfo{year}{2008}).

\bibitem[{\citenamefont{Gogolin and Eisert}(2016)}]{GoEi16}
\bibinfo{author}{\bibfnamefont{C.}~\bibnamefont{Gogolin}} \bibnamefont{and}
  \bibinfo{author}{\bibfnamefont{J.}~\bibnamefont{Eisert}},
  \bibinfo{journal}{Rep. Prog. Phys.} \textbf{\bibinfo{volume}{79}},
  \bibinfo{pages}{056001} (\bibinfo{year}{2016}).

\bibitem[{\citenamefont{Aharonov et~al.}(2020)\citenamefont{Aharonov, Popescu,
  and Rohrlich}}]{AhPoR21}
\bibinfo{author}{\bibfnamefont{Y.}~\bibnamefont{Aharonov}},
  \bibinfo{author}{\bibfnamefont{S.}~\bibnamefont{Popescu}}, \bibnamefont{and}
  \bibinfo{author}{\bibfnamefont{D.}~\bibnamefont{Rohrlich}},
  \bibinfo{journal}{Proceedings of the National Academy of Sciences}
  \textbf{\bibinfo{volume}{118}}, \bibinfo{pages}{e1921529118}
  (\bibinfo{year}{2020}).

\bibitem[{\citenamefont{Lebowitz}(2021)}]{Leb21}
\bibinfo{author}{\bibfnamefont{J.~L.} \bibnamefont{Lebowitz}},
  \bibinfo{journal}{IAMP News Bulletin} pp. \bibinfo{pages}{4--23}
  (\bibinfo{year}{2021}),
  \urlprefix\url{http://www.iamp.org/bulletins/Bulletin-Apr2021-print.pdf}.

\bibitem[{\citenamefont{Lebowitz}(1993)}]{Leb93}
\bibinfo{author}{\bibfnamefont{J.~L.} \bibnamefont{Lebowitz}},
  \bibinfo{journal}{Physics Today} \textbf{\bibinfo{volume}{46}},
  \bibinfo{pages}{(9)32} (\bibinfo{year}{1993}).

\bibitem[{\citenamefont{Choi et~al.}(2016)\citenamefont{Choi, Hild, Zeiher,
  Schau{\ss}, Rubio-Abadal, Yefsah, Khemani, Huse, Bloch, and
  Gross}}]{C+BlGr16}
\bibinfo{author}{\bibfnamefont{J.-Y.} \bibnamefont{Choi}},
  \bibinfo{author}{\bibfnamefont{S.}~\bibnamefont{Hild}},
  \bibinfo{author}{\bibfnamefont{J.}~\bibnamefont{Zeiher}},
  \bibinfo{author}{\bibfnamefont{P.}~\bibnamefont{Schau{\ss}}},
  \bibinfo{author}{\bibfnamefont{A.}~\bibnamefont{Rubio-Abadal}},
  \bibinfo{author}{\bibfnamefont{T.}~\bibnamefont{Yefsah}},
  \bibinfo{author}{\bibfnamefont{V.}~\bibnamefont{Khemani}},
  \bibinfo{author}{\bibfnamefont{D.~A.} \bibnamefont{Huse}},
  \bibinfo{author}{\bibfnamefont{I.}~\bibnamefont{Bloch}}, \bibnamefont{and}
  \bibinfo{author}{\bibfnamefont{C.}~\bibnamefont{Gross}},
  \bibinfo{journal}{Science} \textbf{\bibinfo{volume}{352}},
  \bibinfo{pages}{1547} (\bibinfo{year}{2016}).

\bibitem[{\citenamefont{Monroe et~al.}(2016)\citenamefont{Monroe, Schoelkopf,
  and Lukin}}]{MSLu16}
\bibinfo{author}{\bibfnamefont{C.~R.} \bibnamefont{Monroe}},
  \bibinfo{author}{\bibfnamefont{R.~J.} \bibnamefont{Schoelkopf}},
  \bibnamefont{and} \bibinfo{author}{\bibfnamefont{M.~D.} \bibnamefont{Lukin}},
  \bibinfo{journal}{Sci. Amer.} \textbf{\bibinfo{volume}{314/05}},
  \bibinfo{pages}{50} (\bibinfo{year}{2016}).

\bibitem[{\citenamefont{{Bernien} et~al.}(2017)\citenamefont{{Bernien},
  {Schwartz}, {Keesling}, {Levine}, {Omran}, {Pichler}, {Choi}, {Zibrov},
  {Endres}, {Greiner} et~al.}}]{B+Lu17}
\bibinfo{author}{\bibfnamefont{H.}~\bibnamefont{{Bernien}}},
  \bibinfo{author}{\bibfnamefont{S.}~\bibnamefont{{Schwartz}}},
  \bibinfo{author}{\bibfnamefont{A.}~\bibnamefont{{Keesling}}},
  \bibinfo{author}{\bibfnamefont{H.}~\bibnamefont{{Levine}}},
  \bibinfo{author}{\bibfnamefont{A.}~\bibnamefont{{Omran}}},
  \bibinfo{author}{\bibfnamefont{H.}~\bibnamefont{{Pichler}}},
  \bibinfo{author}{\bibfnamefont{S.}~\bibnamefont{{Choi}}},
  \bibinfo{author}{\bibfnamefont{A.~S.} \bibnamefont{{Zibrov}}},
  \bibinfo{author}{\bibfnamefont{M.}~\bibnamefont{{Endres}}},
  \bibinfo{author}{\bibfnamefont{M.}~\bibnamefont{{Greiner}}},
  \bibnamefont{et~al.}, \bibinfo{journal}{Nature (London)}
  \textbf{\bibinfo{volume}{551}}, \bibinfo{pages}{579} (\bibinfo{year}{2017}).

\bibitem[{\citenamefont{Neill et~al.}(2018)\citenamefont{Neill, Roushan,
  Kechedzhi, Boixo, Isakov, Smelyanskiy, Megrant, Chiaro, Dunsworth, Arya
  et~al.}}]{N+Ma18}
\bibinfo{author}{\bibfnamefont{C.}~\bibnamefont{Neill}},
  \bibinfo{author}{\bibfnamefont{P.}~\bibnamefont{Roushan}},
  \bibinfo{author}{\bibfnamefont{K.}~\bibnamefont{Kechedzhi}},
  \bibinfo{author}{\bibfnamefont{S.}~\bibnamefont{Boixo}},
  \bibinfo{author}{\bibfnamefont{S.~V.} \bibnamefont{Isakov}},
  \bibinfo{author}{\bibfnamefont{V.}~\bibnamefont{Smelyanskiy}},
  \bibinfo{author}{\bibfnamefont{A.}~\bibnamefont{Megrant}},
  \bibinfo{author}{\bibfnamefont{B.}~\bibnamefont{Chiaro}},
  \bibinfo{author}{\bibfnamefont{A.}~\bibnamefont{Dunsworth}},
  \bibinfo{author}{\bibfnamefont{K.}~\bibnamefont{Arya}}, \bibnamefont{et~al.},
  \bibinfo{journal}{Science} \textbf{\bibinfo{volume}{360}},
  \bibinfo{pages}{195} (\bibinfo{year}{2018}).

\bibitem[{\citenamefont{Yao and Nayak}(2018)}]{YaNa18}
\bibinfo{author}{\bibfnamefont{N.~Y.} \bibnamefont{Yao}} \bibnamefont{and}
  \bibinfo{author}{\bibfnamefont{C.}~\bibnamefont{Nayak}},
  \bibinfo{journal}{Physics Today} \textbf{\bibinfo{volume}{71}},
  \bibinfo{pages}{40} (\bibinfo{year}{2018}).

\bibitem[{\citenamefont{Zhou et~al.}(2020)\citenamefont{Zhou, Choi, Choi,
  Landig, Douglas, Isoya, Jelezko, Onoda, Sumiya, Cappellaro
  et~al.}}]{Z+Ca+Lu20}
\bibinfo{author}{\bibfnamefont{H.}~\bibnamefont{Zhou}},
  \bibinfo{author}{\bibfnamefont{J.}~\bibnamefont{Choi}},
  \bibinfo{author}{\bibfnamefont{S.}~\bibnamefont{Choi}},
  \bibinfo{author}{\bibfnamefont{R.}~\bibnamefont{Landig}},
  \bibinfo{author}{\bibfnamefont{A.~M.} \bibnamefont{Douglas}},
  \bibinfo{author}{\bibfnamefont{J.}~\bibnamefont{Isoya}},
  \bibinfo{author}{\bibfnamefont{F.}~\bibnamefont{Jelezko}},
  \bibinfo{author}{\bibfnamefont{S.}~\bibnamefont{Onoda}},
  \bibinfo{author}{\bibfnamefont{H.}~\bibnamefont{Sumiya}},
  \bibinfo{author}{\bibfnamefont{P.}~\bibnamefont{Cappellaro}},
  \bibnamefont{et~al.}, \bibinfo{journal}{Phys. Rev. X}
  \textbf{\bibinfo{volume}{10}}, \bibinfo{pages}{031003}
  (\bibinfo{year}{2020}).

\bibitem[{\citenamefont{Sekino and Susskind}(2008)}]{SeSu08}
\bibinfo{author}{\bibfnamefont{Y.}~\bibnamefont{Sekino}} \bibnamefont{and}
  \bibinfo{author}{\bibfnamefont{L.}~\bibnamefont{Susskind}},
  \bibinfo{journal}{J. High Energy Phys.} \textbf{\bibinfo{volume}{2008}},
  \bibinfo{pages}{065} (\bibinfo{year}{2008}).

\bibitem[{\citenamefont{Maldacena et~al.}(2016)\citenamefont{Maldacena,
  Shenker, and Stanford}}]{MalSSt16}
\bibinfo{author}{\bibfnamefont{J.~M.} \bibnamefont{Maldacena}},
  \bibinfo{author}{\bibfnamefont{S.~H.} \bibnamefont{Shenker}},
  \bibnamefont{and} \bibinfo{author}{\bibfnamefont{D.}~\bibnamefont{Stanford}},
  \bibinfo{journal}{J. High Energy Phys.} \textbf{\bibinfo{volume}{2016}}
  (\bibinfo{year}{2016}).

\bibitem[{\citenamefont{Larkin and Ovchinikov}(1969)}]{LaOv69}
\bibinfo{author}{\bibfnamefont{A.~I.} \bibnamefont{Larkin}} \bibnamefont{and}
  \bibinfo{author}{\bibfnamefont{Y.~N.} \bibnamefont{Ovchinikov}},
  \bibinfo{journal}{Sov. Phys. JETP} \textbf{\bibinfo{volume}{28}},
  \bibinfo{pages}{1200} (\bibinfo{year}{1969}),
  \urlprefix\url{http://www.jetp.ac.ru/cgi-bin/dn/e_028_06_1200.pdf}.

\bibitem[{\citenamefont{Laughlin}(1987)}]{Lau87}
\bibinfo{author}{\bibfnamefont{R.~B.} \bibnamefont{Laughlin}},
  \bibinfo{journal}{Nuclear Physics B - Proceedings Supplements}
  \textbf{\bibinfo{volume}{2}}, \bibinfo{pages}{213} (\bibinfo{year}{1987}).

\bibitem[{\citenamefont{Kitaev}(2017)}]{Ki17}
\bibinfo{author}{\bibfnamefont{A.}~\bibnamefont{Kitaev}}, in
  \emph{\bibinfo{booktitle}{Brown Phys. Colloq., March 6}}
  (\bibinfo{year}{2017}), \urlprefix\url{https://youtu.be/pFBAm7UCFHQ}.

\bibitem[{\citenamefont{Zangara et~al.}(2016)\citenamefont{Zangara, Bendersky,
  Levstein, and Pastawski}}]{Zg+Pa16}
\bibinfo{author}{\bibfnamefont{P.~R.} \bibnamefont{Zangara}},
  \bibinfo{author}{\bibfnamefont{D.}~\bibnamefont{Bendersky}},
  \bibinfo{author}{\bibfnamefont{P.~R.} \bibnamefont{Levstein}},
  \bibnamefont{and} \bibinfo{author}{\bibfnamefont{H.~M.}
  \bibnamefont{Pastawski}}, \bibinfo{journal}{Phil. Trans. R. Soc. A}
  \textbf{\bibinfo{volume}{374}}, \bibinfo{pages}{20150163}
  (\bibinfo{year}{2016}).

\bibitem[{\citenamefont{Kurchan}(2018)}]{Ku18}
\bibinfo{author}{\bibfnamefont{J.}~\bibnamefont{Kurchan}}, \bibinfo{journal}{J.
  Stat. Phys.} \textbf{\bibinfo{volume}{171}}, \bibinfo{pages}{965}
  (\bibinfo{year}{2018}).

\bibitem[{\citenamefont{Schleier-Smith}(2017)}]{SchS17}
\bibinfo{author}{\bibfnamefont{M.}~\bibnamefont{Schleier-Smith}},
  \bibinfo{journal}{Nature Physics} \textbf{\bibinfo{volume}{13}},
  \bibinfo{pages}{724} (\bibinfo{year}{2017}).

\bibitem[{\citenamefont{Yan et~al.}(2020)\citenamefont{Yan, Cincio, and
  Zurek}}]{YCZu20}
\bibinfo{author}{\bibfnamefont{B.}~\bibnamefont{Yan}},
  \bibinfo{author}{\bibfnamefont{L.}~\bibnamefont{Cincio}}, \bibnamefont{and}
  \bibinfo{author}{\bibfnamefont{W.~H.} \bibnamefont{Zurek}},
  \bibinfo{journal}{Phys. Rev. Lett.} \textbf{\bibinfo{volume}{124}},
  \bibinfo{pages}{160603} (\bibinfo{year}{2020}).

\bibitem[{\citenamefont{Levstein et~al.}(1998)\citenamefont{Levstein, Usaj, and
  Pastawski}}]{LUPa98}
\bibinfo{author}{\bibfnamefont{P.~R.} \bibnamefont{Levstein}},
  \bibinfo{author}{\bibfnamefont{G.}~\bibnamefont{Usaj}}, \bibnamefont{and}
  \bibinfo{author}{\bibfnamefont{H.~M.} \bibnamefont{Pastawski}},
  \bibinfo{journal}{J. Chem. Phys.} \textbf{\bibinfo{volume}{108}},
  \bibinfo{pages}{2718} (\bibinfo{year}{1998}).

\bibitem[{\citenamefont{Jalabert and Pastawski}(2001)}]{JaPa01}
\bibinfo{author}{\bibfnamefont{R.~A.} \bibnamefont{Jalabert}} \bibnamefont{and}
  \bibinfo{author}{\bibfnamefont{H.~M.} \bibnamefont{Pastawski}},
  \bibinfo{journal}{Phys. Rev. Lett.} \textbf{\bibinfo{volume}{86}},
  \bibinfo{pages}{2490} (\bibinfo{year}{2001}).

\bibitem[{\citenamefont{Goussev et~al.}(2012)\citenamefont{Goussev, Jalabert,
  Pastawski, and Wisniacki}}]{GJaPaWi12}
\bibinfo{author}{\bibfnamefont{A.}~\bibnamefont{Goussev}},
  \bibinfo{author}{\bibfnamefont{R.~A.} \bibnamefont{Jalabert}},
  \bibinfo{author}{\bibfnamefont{H.~M.} \bibnamefont{Pastawski}},
  \bibnamefont{and} \bibinfo{author}{\bibfnamefont{D.~A.}
  \bibnamefont{Wisniacki}}, \bibinfo{journal}{Scholarpedia}
  \textbf{\bibinfo{volume}{7}}, \bibinfo{pages}{11687} (\bibinfo{year}{2012}).

\bibitem[{\citenamefont{Aleiner et~al.}(2016)\citenamefont{Aleiner, Faoro, and
  Ioffe}}]{AlFIo16}
\bibinfo{author}{\bibfnamefont{I.~L.} \bibnamefont{Aleiner}},
  \bibinfo{author}{\bibfnamefont{L.}~\bibnamefont{Faoro}}, \bibnamefont{and}
  \bibinfo{author}{\bibfnamefont{L.~B.} \bibnamefont{Ioffe}},
  \bibinfo{journal}{Ann. Phys. (New York)} \textbf{\bibinfo{volume}{375}},
  \bibinfo{pages}{378} (\bibinfo{year}{2016}).

\bibitem[{\citenamefont{Rozenbaum et~al.}(2017)\citenamefont{Rozenbaum,
  Ganeshan, and Galitski}}]{RGGa17}
\bibinfo{author}{\bibfnamefont{E.~B.} \bibnamefont{Rozenbaum}},
  \bibinfo{author}{\bibfnamefont{S.}~\bibnamefont{Ganeshan}}, \bibnamefont{and}
  \bibinfo{author}{\bibfnamefont{V.}~\bibnamefont{Galitski}},
  \bibinfo{journal}{Phys. Rev. Lett.} \textbf{\bibinfo{volume}{118}},
  \bibinfo{pages}{086801} (\bibinfo{year}{2017}).

\bibitem[{\citenamefont{Li et~al.}(2017)\citenamefont{Li, Fan, Wang, Ye, Zeng,
  Zhai, Peng, and Du}}]{Li+Du17}
\bibinfo{author}{\bibfnamefont{J.}~\bibnamefont{Li}},
  \bibinfo{author}{\bibfnamefont{R.}~\bibnamefont{Fan}},
  \bibinfo{author}{\bibfnamefont{H.}~\bibnamefont{Wang}},
  \bibinfo{author}{\bibfnamefont{B.}~\bibnamefont{Ye}},
  \bibinfo{author}{\bibfnamefont{B.}~\bibnamefont{Zeng}},
  \bibinfo{author}{\bibfnamefont{H.}~\bibnamefont{Zhai}},
  \bibinfo{author}{\bibfnamefont{X.}~\bibnamefont{Peng}}, \bibnamefont{and}
  \bibinfo{author}{\bibfnamefont{J.}~\bibnamefont{Du}}, \bibinfo{journal}{Phys.
  Rev. X} \textbf{\bibinfo{volume}{7}}, \bibinfo{pages}{031011}
  (\bibinfo{year}{2017}).

\bibitem[{\citenamefont{Lewis-Swan et~al.}(2019)\citenamefont{Lewis-Swan,
  Safavi-Naini, Bollinger, and Rey}}]{Le+Re19}
\bibinfo{author}{\bibfnamefont{R.~J.} \bibnamefont{Lewis-Swan}},
  \bibinfo{author}{\bibfnamefont{A.}~\bibnamefont{Safavi-Naini}},
  \bibinfo{author}{\bibfnamefont{J.~J.} \bibnamefont{Bollinger}},
  \bibnamefont{and} \bibinfo{author}{\bibfnamefont{A.~M.} \bibnamefont{Rey}},
  \bibinfo{journal}{Nat. Commun.} \textbf{\bibinfo{volume}{10}},
  \bibinfo{pages}{1581} (\bibinfo{year}{2019}).

\bibitem[{\citenamefont{Goldfriend and Kurchan}(2020)}]{GoKur20}
\bibinfo{author}{\bibfnamefont{T.}~\bibnamefont{Goldfriend}} \bibnamefont{and}
  \bibinfo{author}{\bibfnamefont{J.}~\bibnamefont{Kurchan}},
  \bibinfo{journal}{Phy. Rev. E} \textbf{\bibinfo{volume}{102}},
  \bibinfo{pages}{022201} (\bibinfo{year}{2020}).

\bibitem[{\citenamefont{Maldacena}(2020)}]{Mald20}
\bibinfo{author}{\bibfnamefont{J.~M.} \bibnamefont{Maldacena}},
  \bibinfo{journal}{Nature Reviews Physics} \textbf{\bibinfo{volume}{2}},
  \bibinfo{pages}{123} (\bibinfo{year}{2020}).

\bibitem[{\citenamefont{Horgan}(2020{\natexlab{a}})}]{Hor20}
\bibinfo{author}{\bibfnamefont{J.}~\bibnamefont{Horgan}},
  \bibinfo{journal}{Scientific American}  (\bibinfo{year}{2020}{\natexlab{a}}),
  \urlprefix\url{https://www.scientificamerican.com/article/will-the-universe-remember-us-after-were-gone/}.

\bibitem[{\citenamefont{Anderson}(1978)}]{An78}
\bibinfo{author}{\bibfnamefont{P.~W.} \bibnamefont{Anderson}},
  \bibinfo{journal}{Rev. Mod. Phys.} \textbf{\bibinfo{volume}{50}},
  \bibinfo{pages}{191} (\bibinfo{year}{1978}).

\bibitem[{\citenamefont{Anderson}(1972)}]{An72}
\bibinfo{author}{\bibfnamefont{P.~W.} \bibnamefont{Anderson}},
  \bibinfo{journal}{Science} \textbf{\bibinfo{volume}{177}},
  \bibinfo{pages}{393} (\bibinfo{year}{1972}).

\bibitem[{\citenamefont{Zangara and Pastawski}(2017)}]{ZgPa17}
\bibinfo{author}{\bibfnamefont{P.~R.} \bibnamefont{Zangara}} \bibnamefont{and}
  \bibinfo{author}{\bibfnamefont{H.~M.} \bibnamefont{Pastawski}},
  \bibinfo{journal}{Phys. Scr.} \textbf{\bibinfo{volume}{92}},
  \bibinfo{pages}{033001} (\bibinfo{year}{2017}).

\bibitem[{\citenamefont{Ghirardi et~al.}(1986)\citenamefont{Ghirardi, Rimini,
  and Weber}}]{GRW86}
\bibinfo{author}{\bibfnamefont{G.~C.} \bibnamefont{Ghirardi}},
  \bibinfo{author}{\bibfnamefont{A.}~\bibnamefont{Rimini}}, \bibnamefont{and}
  \bibinfo{author}{\bibfnamefont{T.}~\bibnamefont{Weber}},
  \bibinfo{journal}{Phys. Rev. D} \textbf{\bibinfo{volume}{34}},
  \bibinfo{pages}{470} (\bibinfo{year}{1986}).

\bibitem[{\citenamefont{Weinberg}(2012)}]{Wein12}
\bibinfo{author}{\bibfnamefont{S.}~\bibnamefont{Weinberg}},
  \bibinfo{journal}{Phys. Rev. A} \textbf{\bibinfo{volume}{85}},
  \bibinfo{pages}{062116} (\bibinfo{year}{2012}).

\bibitem[{\citenamefont{Feher}(1959)}]{Feher1959}
\bibinfo{author}{\bibfnamefont{G.}~\bibnamefont{Feher}},
  \bibinfo{journal}{Phys. Rev.} \textbf{\bibinfo{volume}{114}},
  \bibinfo{pages}{1219} (\bibinfo{year}{1959}).

\bibitem[{\citenamefont{Sachdev}(2009)}]{Sachdev2009}
\bibinfo{author}{\bibfnamefont{S.}~\bibnamefont{Sachdev}},
  \emph{\bibinfo{title}{Quantum Phase Transitions}}
  (\bibinfo{publisher}{Cambridge University Press}, \bibinfo{year}{2009}).

\bibitem[{\citenamefont{Anderson}(1954)}]{An54}
\bibinfo{author}{\bibfnamefont{P.~W.} \bibnamefont{Anderson}},
  \bibinfo{journal}{J. Phys. Soc. Jpn.} \textbf{\bibinfo{volume}{9}},
  \bibinfo{pages}{316} (\bibinfo{year}{1954}).

\bibitem[{\citenamefont{Quan et~al.}(2006)\citenamefont{Quan, Song, Liu,
  Zanardi, and Sun}}]{Qu+ZaSu06}
\bibinfo{author}{\bibfnamefont{H.~T.} \bibnamefont{Quan}},
  \bibinfo{author}{\bibfnamefont{Z.}~\bibnamefont{Song}},
  \bibinfo{author}{\bibfnamefont{X.~F.} \bibnamefont{Liu}},
  \bibinfo{author}{\bibfnamefont{P.}~\bibnamefont{Zanardi}}, \bibnamefont{and}
  \bibinfo{author}{\bibfnamefont{C.~P.} \bibnamefont{Sun}},
  \bibinfo{journal}{Phys. Rev. Lett.} \textbf{\bibinfo{volume}{96}},
  \bibinfo{pages}{140604} (\bibinfo{year}{2006}).

\bibitem[{\citenamefont{Zhang et~al.}(2008)\citenamefont{Zhang, Peng,
  Rajendran, and Suter}}]{Z+Su08}
\bibinfo{author}{\bibfnamefont{J.}~\bibnamefont{Zhang}},
  \bibinfo{author}{\bibfnamefont{X.}~\bibnamefont{Peng}},
  \bibinfo{author}{\bibfnamefont{N.}~\bibnamefont{Rajendran}},
  \bibnamefont{and} \bibinfo{author}{\bibfnamefont{D.}~\bibnamefont{Suter}},
  \bibinfo{journal}{Phys. Rev. Lett.} \textbf{\bibinfo{volume}{100}},
  \bibinfo{pages}{100501} (\bibinfo{year}{2008}).

\bibitem[{\citenamefont{\'Alvarez et~al.}(2015)\citenamefont{\'Alvarez, Suter,
  and Kaiser}}]{ASuK15}
\bibinfo{author}{\bibfnamefont{G.~A.} \bibnamefont{\'Alvarez}},
  \bibinfo{author}{\bibfnamefont{D.}~\bibnamefont{Suter}}, \bibnamefont{and}
  \bibinfo{author}{\bibfnamefont{R.}~\bibnamefont{Kaiser}},
  \bibinfo{journal}{Science} \textbf{\bibinfo{volume}{349}},
  \bibinfo{pages}{846} (\bibinfo{year}{2015}).

\bibitem[{\citenamefont{Wei et~al.}(2018)\citenamefont{Wei, Ramanathan, and
  Cappellaro}}]{WeiChCa18}
\bibinfo{author}{\bibfnamefont{K.~X.} \bibnamefont{Wei}},
  \bibinfo{author}{\bibfnamefont{C.}~\bibnamefont{Ramanathan}},
  \bibnamefont{and}
  \bibinfo{author}{\bibfnamefont{P.}~\bibnamefont{Cappellaro}},
  \bibinfo{journal}{Phys. Rev. Lett.} \textbf{\bibinfo{volume}{120}},
  \bibinfo{pages}{070501} (\bibinfo{year}{2018}).

\bibitem[{\citenamefont{Maurer et~al.}(2012)\citenamefont{Maurer, Kucsko,
  Latta, Jiang, Yao, Bennett, Pastawski, Hunger, Chisholm, Markham
  et~al.}}]{M+CiLu12}
\bibinfo{author}{\bibfnamefont{P.~C.} \bibnamefont{Maurer}},
  \bibinfo{author}{\bibfnamefont{G.}~\bibnamefont{Kucsko}},
  \bibinfo{author}{\bibfnamefont{C.}~\bibnamefont{Latta}},
  \bibinfo{author}{\bibfnamefont{L.}~\bibnamefont{Jiang}},
  \bibinfo{author}{\bibfnamefont{N.~Y.} \bibnamefont{Yao}},
  \bibinfo{author}{\bibfnamefont{S.~D.} \bibnamefont{Bennett}},
  \bibinfo{author}{\bibfnamefont{F.}~\bibnamefont{Pastawski}},
  \bibinfo{author}{\bibfnamefont{D.}~\bibnamefont{Hunger}},
  \bibinfo{author}{\bibfnamefont{N.}~\bibnamefont{Chisholm}},
  \bibinfo{author}{\bibfnamefont{M.}~\bibnamefont{Markham}},
  \bibnamefont{et~al.}, \bibinfo{journal}{Science}
  \textbf{\bibinfo{volume}{336}}, \bibinfo{pages}{1283} (\bibinfo{year}{2012}).

\bibitem[{\citenamefont{Zhang and Cory}(1998)}]{ZCo98}
\bibinfo{author}{\bibfnamefont{W.}~\bibnamefont{Zhang}} \bibnamefont{and}
  \bibinfo{author}{\bibfnamefont{D.~G.} \bibnamefont{Cory}},
  \bibinfo{journal}{Phys. Rev. Lett.} \textbf{\bibinfo{volume}{80}},
  \bibinfo{pages}{1324} (\bibinfo{year}{1998}).

\bibitem[{\citenamefont{Boutis et~al.}(2004)\citenamefont{Boutis, Greenbaum,
  Cho, Cory, and Ramanathan}}]{BoCo+04}
\bibinfo{author}{\bibfnamefont{G.~S.} \bibnamefont{Boutis}},
  \bibinfo{author}{\bibfnamefont{D.}~\bibnamefont{Greenbaum}},
  \bibinfo{author}{\bibfnamefont{H.}~\bibnamefont{Cho}},
  \bibinfo{author}{\bibfnamefont{D.~G.} \bibnamefont{Cory}}, \bibnamefont{and}
  \bibinfo{author}{\bibfnamefont{C.}~\bibnamefont{Ramanathan}},
  \bibinfo{journal}{Phys. Rev. Lett.} \textbf{\bibinfo{volume}{92}},
  \bibinfo{pages}{137201} (\bibinfo{year}{2004}).

\bibitem[{\citenamefont{Hahn}(1950)}]{Hh50}
\bibinfo{author}{\bibfnamefont{E.}~\bibnamefont{Hahn}}, \bibinfo{journal}{Phys.
  Rev.} \textbf{\bibinfo{volume}{80}}, \bibinfo{pages}{580}
  (\bibinfo{year}{1950}).

\bibitem[{\citenamefont{Brewer and Hahn}(1984)}]{BrHh84}
\bibinfo{author}{\bibfnamefont{R.~G.} \bibnamefont{Brewer}} \bibnamefont{and}
  \bibinfo{author}{\bibfnamefont{E.~L.} \bibnamefont{Hahn}},
  \bibinfo{journal}{Scientific American} \textbf{\bibinfo{volume}{251}},
  \bibinfo{pages}{(6)50} (\bibinfo{year}{1984}),
  \urlprefix\url{http://www.jstor.org/stable/24969498}.

\bibitem[{\citenamefont{Rhim et~al.}(1971)\citenamefont{Rhim, Pines, and
  Waugh}}]{RPiWa71}
\bibinfo{author}{\bibfnamefont{W.-K.} \bibnamefont{Rhim}},
  \bibinfo{author}{\bibfnamefont{A.}~\bibnamefont{Pines}}, \bibnamefont{and}
  \bibinfo{author}{\bibfnamefont{J.~S.} \bibnamefont{Waugh}},
  \bibinfo{journal}{Phys. Rev. B} \textbf{\bibinfo{volume}{3}},
  \bibinfo{pages}{684} (\bibinfo{year}{1971}).

\bibitem[{\citenamefont{Munowitz and Pines}(1986)}]{MuPi86}
\bibinfo{author}{\bibfnamefont{M.}~\bibnamefont{Munowitz}} \bibnamefont{and}
  \bibinfo{author}{\bibfnamefont{A.}~\bibnamefont{Pines}},
  \bibinfo{journal}{Science} \textbf{\bibinfo{volume}{233}},
  \bibinfo{pages}{525} (\bibinfo{year}{1986}).

\bibitem[{\citenamefont{Niknam et~al.}(2020)\citenamefont{Niknam, Santos, and
  Cory}}]{NSCo20}
\bibinfo{author}{\bibfnamefont{M.}~\bibnamefont{Niknam}},
  \bibinfo{author}{\bibfnamefont{L.~F.} \bibnamefont{Santos}},
  \bibnamefont{and} \bibinfo{author}{\bibfnamefont{D.~G.} \bibnamefont{Cory}},
  \bibinfo{journal}{Phys. Rev. Research} \textbf{\bibinfo{volume}{2}},
  \bibinfo{pages}{013200} (\bibinfo{year}{2020}).

\bibitem[{\citenamefont{S\'anchez et~al.}(2020)\citenamefont{S\'anchez,
  Chattah, Wei, Buljubasich, Cappellaro, and Pastawski}}]{Sa+Pa20}
\bibinfo{author}{\bibfnamefont{C.~M.} \bibnamefont{S\'anchez}},
  \bibinfo{author}{\bibfnamefont{A.~K.} \bibnamefont{Chattah}},
  \bibinfo{author}{\bibfnamefont{K.~X.} \bibnamefont{Wei}},
  \bibinfo{author}{\bibfnamefont{L.}~\bibnamefont{Buljubasich}},
  \bibinfo{author}{\bibfnamefont{P.}~\bibnamefont{Cappellaro}},
  \bibnamefont{and} \bibinfo{author}{\bibfnamefont{H.~M.}
  \bibnamefont{Pastawski}}, \bibinfo{journal}{Phys. Rev. Lett.}
  \textbf{\bibinfo{volume}{124}}, \bibinfo{pages}{030601}
  (\bibinfo{year}{2020}).

\bibitem[{\citenamefont{Zhang et~al.}(1992)\citenamefont{Zhang, Meier, and
  Ernst}}]{ZMEr92}
\bibinfo{author}{\bibfnamefont{S.}~\bibnamefont{Zhang}},
  \bibinfo{author}{\bibfnamefont{B.~H.} \bibnamefont{Meier}}, \bibnamefont{and}
  \bibinfo{author}{\bibfnamefont{R.~R.} \bibnamefont{Ernst}},
  \bibinfo{journal}{Phys.Rev.Lett.} \textbf{\bibinfo{volume}{69}},
  \bibinfo{pages}{2149} (\bibinfo{year}{1992}).

\bibitem[{\citenamefont{Usaj et~al.}(1998)\citenamefont{Usaj, Pastawski, and
  Levstein}}]{UPaLe98}
\bibinfo{author}{\bibfnamefont{G.}~\bibnamefont{Usaj}},
  \bibinfo{author}{\bibfnamefont{H.~M.} \bibnamefont{Pastawski}},
  \bibnamefont{and} \bibinfo{author}{\bibfnamefont{P.~R.}
  \bibnamefont{Levstein}}, \bibinfo{journal}{Mol. Phys.}
  \textbf{\bibinfo{volume}{95}}, \bibinfo{pages}{1229} (\bibinfo{year}{1998}).

\bibitem[{\citenamefont{Levstein et~al.}(2004)\citenamefont{Levstein, Chattah,
  Pastawski, Raya, and Hirschinger}}]{LeChPa04}
\bibinfo{author}{\bibfnamefont{P.~R.} \bibnamefont{Levstein}},
  \bibinfo{author}{\bibfnamefont{A.~K.} \bibnamefont{Chattah}},
  \bibinfo{author}{\bibfnamefont{H.~M.} \bibnamefont{Pastawski}},
  \bibinfo{author}{\bibfnamefont{J.}~\bibnamefont{Raya}}, \bibnamefont{and}
  \bibinfo{author}{\bibfnamefont{J.}~\bibnamefont{Hirschinger}},
  \bibinfo{journal}{J. Chem. Phys.} \textbf{\bibinfo{volume}{121}},
  \bibinfo{pages}{7313} (\bibinfo{year}{2004}).

\bibitem[{\citenamefont{Ernst}(1996)}]{ErPriv96}
\bibinfo{author}{\bibfnamefont{R.~R.} \bibnamefont{Ernst}},
  \bibinfo{journal}{private communication}  (\bibinfo{year}{1996}).

\bibitem[{\citenamefont{Cory}(1997)}]{CoPriv97}
\bibinfo{author}{\bibfnamefont{D.~G.} \bibnamefont{Cory}},
  \bibinfo{journal}{private comm.}  (\bibinfo{year}{1997}).

\bibitem[{\citenamefont{Pastawski et~al.}(2000)\citenamefont{Pastawski,
  Levstein, Usaj, Raya, and Hirschinger}}]{P00}
\bibinfo{author}{\bibfnamefont{H.~M.} \bibnamefont{Pastawski}},
  \bibinfo{author}{\bibfnamefont{P.}~\bibnamefont{Levstein}},
  \bibinfo{author}{\bibfnamefont{G.}~\bibnamefont{Usaj}},
  \bibinfo{author}{\bibfnamefont{J.}~\bibnamefont{Raya}}, \bibnamefont{and}
  \bibinfo{author}{\bibfnamefont{J.}~\bibnamefont{Hirschinger}},
  \bibinfo{journal}{Physica (Utrecht)} \textbf{\bibinfo{volume}{283A}},
  \bibinfo{pages}{166} (\bibinfo{year}{2000}).

\bibitem[{\citenamefont{Morgan et~al.}(2012)\citenamefont{Morgan, Oganesyan,
  and Boutis}}]{MOgBo12}
\bibinfo{author}{\bibfnamefont{S.~W.} \bibnamefont{Morgan}},
  \bibinfo{author}{\bibfnamefont{V.}~\bibnamefont{Oganesyan}},
  \bibnamefont{and} \bibinfo{author}{\bibfnamefont{G.~S.}
  \bibnamefont{Boutis}}, \bibinfo{journal}{Phys. Rev. B}
  \textbf{\bibinfo{volume}{86}}, \bibinfo{pages}{214410}
  (\bibinfo{year}{2012}).

\bibitem[{\citenamefont{S\'anchez et~al.}(2017)\citenamefont{S\'anchez,
  Buljubasich, Pastawski, and Chattah}}]{SaBPC17}
\bibinfo{author}{\bibfnamefont{C.~M.} \bibnamefont{S\'anchez}},
  \bibinfo{author}{\bibfnamefont{L.}~\bibnamefont{Buljubasich}},
  \bibinfo{author}{\bibfnamefont{H.~M.} \bibnamefont{Pastawski}},
  \bibnamefont{and} \bibinfo{author}{\bibfnamefont{A.~K.}
  \bibnamefont{Chattah}}, \bibinfo{journal}{Journal of Magnetic Resonance}
  \textbf{\bibinfo{volume}{281}}, \bibinfo{pages}{75} (\bibinfo{year}{2017}).

\bibitem[{\citenamefont{Rhim et~al.}(1970)\citenamefont{Rhim, Pines, and
  Waugh}}]{RPiWa70}
\bibinfo{author}{\bibfnamefont{W.-K.} \bibnamefont{Rhim}},
  \bibinfo{author}{\bibfnamefont{A.}~\bibnamefont{Pines}}, \bibnamefont{and}
  \bibinfo{author}{\bibfnamefont{J.~S.} \bibnamefont{Waugh}},
  \bibinfo{journal}{Phys. Rev. Lett.} \textbf{\bibinfo{volume}{25}},
  \bibinfo{pages}{218} (\bibinfo{year}{1970}).

\bibitem[{\citenamefont{Haeberlen}(1976)}]{Hb76}
\bibinfo{author}{\bibfnamefont{U.}~\bibnamefont{Haeberlen}},
  \emph{\bibinfo{title}{High Resolution NMR in solids}}
  (\bibinfo{publisher}{Academic Press}, \bibinfo{year}{1976}), ISBN
  \bibinfo{isbn}{0-12-025561-8}.

\bibitem[{\citenamefont{Kuwahara et~al.}(2016)\citenamefont{Kuwahara, Mori, and
  Saito}}]{KuMoSa16}
\bibinfo{author}{\bibfnamefont{T.}~\bibnamefont{Kuwahara}},
  \bibinfo{author}{\bibfnamefont{T.}~\bibnamefont{Mori}}, \bibnamefont{and}
  \bibinfo{author}{\bibfnamefont{K.}~\bibnamefont{Saito}},
  \bibinfo{journal}{Annals of Physics} \textbf{\bibinfo{volume}{367}},
  \bibinfo{pages}{96} (\bibinfo{year}{2016}).

\bibitem[{\citenamefont{Buljubasich et~al.}(2015)\citenamefont{Buljubasich,
  S{\'a}nchez, Dente, Levstein, Chattah, and Pastawski}}]{BuSa+15}
\bibinfo{author}{\bibfnamefont{L.}~\bibnamefont{Buljubasich}},
  \bibinfo{author}{\bibfnamefont{C.~M.} \bibnamefont{S{\'a}nchez}},
  \bibinfo{author}{\bibfnamefont{A.~D.} \bibnamefont{Dente}},
  \bibinfo{author}{\bibfnamefont{P.~R.} \bibnamefont{Levstein}},
  \bibinfo{author}{\bibfnamefont{A.~K.} \bibnamefont{Chattah}},
  \bibnamefont{and} \bibinfo{author}{\bibfnamefont{H.~M.}
  \bibnamefont{Pastawski}}, \bibinfo{journal}{J. Chem. Phys.}
  \textbf{\bibinfo{volume}{143}}, \bibinfo{pages}{164308}
  (\bibinfo{year}{2015}).

\bibitem[{\citenamefont{Slichter}(1990)}]{Sl90}
\bibinfo{author}{\bibfnamefont{C.~P.} \bibnamefont{Slichter}},
  \emph{\bibinfo{title}{Principles of magnetic resonance}}
  (\bibinfo{publisher}{Springer-Verlag}, \bibinfo{address}{Berlin; New York},
  \bibinfo{year}{1990}).

\bibitem[{\citenamefont{Ernst et~al.}(1987)\citenamefont{Ernst, Bodenhausen,
  and Wokaun}}]{Er87}
\bibinfo{author}{\bibfnamefont{R.~R.} \bibnamefont{Ernst}},
  \bibinfo{author}{\bibfnamefont{G.}~\bibnamefont{Bodenhausen}},
  \bibnamefont{and} \bibinfo{author}{\bibfnamefont{A.}~\bibnamefont{Wokaun}},
  \emph{\bibinfo{title}{Principles of nuclear magnetic resonance in one and two
  dimensions}} (\bibinfo{publisher}{Oxford Univ. Press},
  \bibinfo{address}{Oxford}, \bibinfo{year}{1987}).

\bibitem[{\citenamefont{S\'anchez et~al.}(2016)\citenamefont{S\'anchez,
  Levstein, Buljubasich, Pastawski, and Chattah}}]{Sa+Ch16}
\bibinfo{author}{\bibfnamefont{C.~M.} \bibnamefont{S\'anchez}},
  \bibinfo{author}{\bibfnamefont{P.~R.} \bibnamefont{Levstein}},
  \bibinfo{author}{\bibfnamefont{L.}~\bibnamefont{Buljubasich}},
  \bibinfo{author}{\bibfnamefont{H.~M.} \bibnamefont{Pastawski}},
  \bibnamefont{and} \bibinfo{author}{\bibfnamefont{A.~K.}
  \bibnamefont{Chattah}}, \bibinfo{journal}{Phil. Trans. R. Soc. A}
  \textbf{\bibinfo{volume}{374}}, \bibinfo{pages}{20150155}
  (\bibinfo{year}{2016}).

\bibitem[{\citenamefont{Abragam}(2007)}]{Ab61}
\bibinfo{author}{\bibfnamefont{A.}~\bibnamefont{Abragam}},
  \emph{\bibinfo{title}{Principles of Nuclear Magnetism}}
  (\bibinfo{publisher}{Oxford Univ. Press}, \bibinfo{year}{2007}),
  \bibinfo{edition}{reprinted} ed.

\bibitem[{\citenamefont{Yan and Zurek}(2021)}]{YaZu21}
\bibinfo{author}{\bibfnamefont{B.}~\bibnamefont{Yan}} \bibnamefont{and}
  \bibinfo{author}{\bibfnamefont{W.~H.} \bibnamefont{Zurek}}
  (\bibinfo{year}{2021}), \eprint{hep-th/2110.15306}.

\bibitem[{\citenamefont{Flambaum and Izrailev}(2001)}]{FIz01}
\bibinfo{author}{\bibfnamefont{V.~V.} \bibnamefont{Flambaum}} \bibnamefont{and}
  \bibinfo{author}{\bibfnamefont{F.~M.} \bibnamefont{Izrailev}},
  \bibinfo{journal}{Phys. Rev. E} \textbf{\bibinfo{volume}{64}},
  \bibinfo{pages}{026124} (\bibinfo{year}{2001}).

\bibitem[{\citenamefont{Zurek et~al.}(2007)\citenamefont{Zurek, Cucchietti, and
  Paz}}]{ZuCP07}
\bibinfo{author}{\bibfnamefont{W.~H.} \bibnamefont{Zurek}},
  \bibinfo{author}{\bibfnamefont{F.~M.} \bibnamefont{Cucchietti}},
  \bibnamefont{and} \bibinfo{author}{\bibfnamefont{J.~P.} \bibnamefont{Paz}},
  \bibinfo{journal}{Acta Phys. Pol. B} \textbf{\bibinfo{volume}{38}},
  \bibinfo{pages}{1685} (\bibinfo{year}{2007}),
  \urlprefix\url{http://www.actaphys.uj.edu.pl/fulltext?series=Reg&vol=38&page=1685}.

\bibitem[{\citenamefont{{\'{A}}lvarez et~al.}(2006)\citenamefont{{\'{A}}lvarez,
  Danieli, Levstein, and Pastawski}}]{ADLP06}
\bibinfo{author}{\bibfnamefont{G.~A.} \bibnamefont{{\'{A}}lvarez}},
  \bibinfo{author}{\bibfnamefont{E.~P.} \bibnamefont{Danieli}},
  \bibinfo{author}{\bibfnamefont{P.~R.} \bibnamefont{Levstein}},
  \bibnamefont{and} \bibinfo{author}{\bibfnamefont{H.~M.}
  \bibnamefont{Pastawski}}, \bibinfo{journal}{J. Chem. Phys.}
  \textbf{\bibinfo{volume}{124}}, \bibinfo{pages}{194507}
  (\bibinfo{year}{2006}).

\bibitem[{\citenamefont{Cory et~al.}(1997)\citenamefont{Cory, Fahmy, and
  Havel}}]{CoFH97}
\bibinfo{author}{\bibfnamefont{D.~G.} \bibnamefont{Cory}},
  \bibinfo{author}{\bibfnamefont{A.~F.} \bibnamefont{Fahmy}}, \bibnamefont{and}
  \bibinfo{author}{\bibfnamefont{T.~F.} \bibnamefont{Havel}},
  \bibinfo{journal}{PNAS} \textbf{\bibinfo{volume}{94}}, \bibinfo{pages}{1634}
  (\bibinfo{year}{1997}).

\bibitem[{\citenamefont{Pastawski
  et~al.}(1995{\natexlab{a}})\citenamefont{Pastawski, Levstein, and
  Usaj}}]{PLU95}
\bibinfo{author}{\bibfnamefont{H.~M.} \bibnamefont{Pastawski}},
  \bibinfo{author}{\bibfnamefont{P.~R.} \bibnamefont{Levstein}},
  \bibnamefont{and} \bibinfo{author}{\bibfnamefont{G.}~\bibnamefont{Usaj}},
  \bibinfo{journal}{Phys. Rev. Lett.} \textbf{\bibinfo{volume}{75}},
  \bibinfo{pages}{4310} (\bibinfo{year}{1995}{\natexlab{a}}).

\bibitem[{\citenamefont{Cho et~al.}(2005)\citenamefont{Cho, Ladd, Baugh, Cory,
  and Ramanathan}}]{C+CoRa05}
\bibinfo{author}{\bibfnamefont{H.}~\bibnamefont{Cho}},
  \bibinfo{author}{\bibfnamefont{T.~D.} \bibnamefont{Ladd}},
  \bibinfo{author}{\bibfnamefont{J.}~\bibnamefont{Baugh}},
  \bibinfo{author}{\bibfnamefont{D.~G.} \bibnamefont{Cory}}, \bibnamefont{and}
  \bibinfo{author}{\bibfnamefont{C.}~\bibnamefont{Ramanathan}},
  \bibinfo{journal}{Phys. Rev. B} \textbf{\bibinfo{volume}{72}},
  \bibinfo{pages}{054427} (\bibinfo{year}{2005}).

\bibitem[{\citenamefont{Alvarez et~al.}(2010)\citenamefont{Alvarez, Danieli,
  Levstein, and Pastawski}}]{ADLPa10}
\bibinfo{author}{\bibfnamefont{G.~A.} \bibnamefont{Alvarez}},
  \bibinfo{author}{\bibfnamefont{E.~P.} \bibnamefont{Danieli}},
  \bibinfo{author}{\bibfnamefont{P.~R.} \bibnamefont{Levstein}},
  \bibnamefont{and} \bibinfo{author}{\bibfnamefont{H.~M.}
  \bibnamefont{Pastawski}}, \bibinfo{journal}{Phys. Rev. A}
  \textbf{\bibinfo{volume}{82}}, \bibinfo{pages}{012310}
  (\bibinfo{year}{2010}).

\bibitem[{\citenamefont{Bendersky}(2016)}]{Bend16}
\bibinfo{author}{\bibfnamefont{D.}~\bibnamefont{Bendersky}},
  \bibinfo{journal}{Doct. Thesis, Univ. Nac. Córdoba}  (\bibinfo{year}{2016}),
  \urlprefix\url{http://hdl.handle.net/11086/3374}.

\bibitem[{\citenamefont{Swingle}(2018)}]{Sw18}
\bibinfo{author}{\bibfnamefont{B.}~\bibnamefont{Swingle}},
  \bibinfo{journal}{Nat. Phys.} \textbf{\bibinfo{volume}{14}},
  \bibinfo{pages}{988} (\bibinfo{year}{2018}).

\bibitem[{\citenamefont{Lozano-Negro et~al.}(2021)\citenamefont{Lozano-Negro,
  Zangara, and Pastawski}}]{LoZgPa21}
\bibinfo{author}{\bibfnamefont{F.}~\bibnamefont{Lozano-Negro}},
  \bibinfo{author}{\bibfnamefont{P.~R.} \bibnamefont{Zangara}},
  \bibnamefont{and} \bibinfo{author}{\bibfnamefont{H.~M.}
  \bibnamefont{Pastawski}}, \bibinfo{journal}{Chaos, Solitons and Fractals}
  (\bibinfo{year}{2021}), \eprint{2106.07370}.

\bibitem[{\citenamefont{Khitrin}(1997)}]{Kh97}
\bibinfo{author}{\bibfnamefont{A.~K.} \bibnamefont{Khitrin}},
  \bibinfo{journal}{Chem. Phys. Lett.} \textbf{\bibinfo{volume}{274}},
  \bibinfo{pages}{217} (\bibinfo{year}{1997}).

\bibitem[{\citenamefont{Munowitz et~al.}(1987)\citenamefont{Munowitz, Pines,
  and Mehring}}]{MuPiM87}
\bibinfo{author}{\bibfnamefont{M.}~\bibnamefont{Munowitz}},
  \bibinfo{author}{\bibfnamefont{A.}~\bibnamefont{Pines}}, \bibnamefont{and}
  \bibinfo{author}{\bibfnamefont{M.}~\bibnamefont{Mehring}},
  \bibinfo{journal}{J. Chem. Phys.} \textbf{\bibinfo{volume}{86}},
  \bibinfo{pages}{3172} (\bibinfo{year}{1987}).

\bibitem[{\citenamefont{Yoshida and Yao}(2019)}]{YoYa19}
\bibinfo{author}{\bibfnamefont{B.}~\bibnamefont{Yoshida}} \bibnamefont{and}
  \bibinfo{author}{\bibfnamefont{N.~Y.} \bibnamefont{Yao}},
  \bibinfo{journal}{Phys. Rev. X} \textbf{\bibinfo{volume}{9}},
  \bibinfo{pages}{011006} (\bibinfo{year}{2019}).

\bibitem[{\citenamefont{Zobov and Lundin}(2021)}]{ZoLu21}
\bibinfo{author}{\bibfnamefont{V.~E.} \bibnamefont{Zobov}} \bibnamefont{and}
  \bibinfo{author}{\bibfnamefont{A.~A.} \bibnamefont{Lundin}},
  \bibinfo{journal}{Appl. Magn. Reson.}  (\bibinfo{year}{2021}).

\bibitem[{\citenamefont{Aleiner and Larkin}(1996)}]{ALar96}
\bibinfo{author}{\bibfnamefont{I.~L.} \bibnamefont{Aleiner}} \bibnamefont{and}
  \bibinfo{author}{\bibfnamefont{A.~I.} \bibnamefont{Larkin}},
  \bibinfo{journal}{Phys. Rev. B} \textbf{\bibinfo{volume}{54}},
  \bibinfo{pages}{14423} (\bibinfo{year}{1996}).

\bibitem[{\citenamefont{Cucchietti et~al.}(2004)\citenamefont{Cucchietti,
  Pastawski, and Jalabert}}]{CuPJa04}
\bibinfo{author}{\bibfnamefont{F.~M.} \bibnamefont{Cucchietti}},
  \bibinfo{author}{\bibfnamefont{H.~M.} \bibnamefont{Pastawski}},
  \bibnamefont{and} \bibinfo{author}{\bibfnamefont{R.~A.}
  \bibnamefont{Jalabert}}, \bibinfo{journal}{Phys. Rev. B}
  \textbf{\bibinfo{volume}{70}}, \bibinfo{pages}{035311}
  (\bibinfo{year}{2004}), \bibinfo{note}{(see Figs. 4, 5 and 6)}.

\bibitem[{\citenamefont{Casati et~al.}(1987)\citenamefont{Casati, Chirikov,
  Shepelyansky, and Guarnieri}}]{CaChi87}
\bibinfo{author}{\bibfnamefont{G.}~\bibnamefont{Casati}},
  \bibinfo{author}{\bibfnamefont{B.}~\bibnamefont{Chirikov}},
  \bibinfo{author}{\bibfnamefont{D.}~\bibnamefont{Shepelyansky}},
  \bibnamefont{and}
  \bibinfo{author}{\bibfnamefont{I.}~\bibnamefont{Guarnieri}},
  \bibinfo{journal}{Phys. Rep.} \textbf{\bibinfo{volume}{154}},
  \bibinfo{pages}{77} (\bibinfo{year}{1987}).

\bibitem[{\citenamefont{Elsayed and Fine}(2015)}]{ElFi15}
\bibinfo{author}{\bibfnamefont{T.~A.} \bibnamefont{Elsayed}} \bibnamefont{and}
  \bibinfo{author}{\bibfnamefont{B.~V.} \bibnamefont{Fine}},
  \bibinfo{journal}{Phys. Scr.} \textbf{\bibinfo{volume}{T165}},
  \bibinfo{pages}{014011} (\bibinfo{year}{2015}).

\bibitem[{\citenamefont{Jacquod et~al.}(2001)\citenamefont{Jacquod, Silvestrov,
  and Beenakker}}]{JacSiBee01}
\bibinfo{author}{\bibfnamefont{P.}~\bibnamefont{Jacquod}},
  \bibinfo{author}{\bibfnamefont{P.~G.} \bibnamefont{Silvestrov}},
  \bibnamefont{and} \bibinfo{author}{\bibfnamefont{C.~W.~J.}
  \bibnamefont{Beenakker}}, \bibinfo{journal}{Phys. Rev. E}
  \textbf{\bibinfo{volume}{64}}, \bibinfo{pages}{055203(R)}
  (\bibinfo{year}{2001}).

\bibitem[{\citenamefont{Jacquod et~al.}(2002)\citenamefont{Jacquod, Adagideli,
  and Beenakker}}]{JacAdBee02}
\bibinfo{author}{\bibfnamefont{P.}~\bibnamefont{Jacquod}},
  \bibinfo{author}{\bibfnamefont{I.}~\bibnamefont{Adagideli}},
  \bibnamefont{and} \bibinfo{author}{\bibfnamefont{C.~W.~J.}
  \bibnamefont{Beenakker}}, \bibinfo{journal}{Phys. Rev. Lett.}
  \textbf{\bibinfo{volume}{89}}, \bibinfo{pages}{154103}
  (\bibinfo{year}{2002}).

\bibitem[{\citenamefont{van Ede van~der Pals and Gaspard}(1994)}]{vEvdPG94}
\bibinfo{author}{\bibfnamefont{P.}~\bibnamefont{van Ede van~der Pals}}
  \bibnamefont{and} \bibinfo{author}{\bibfnamefont{P.}~\bibnamefont{Gaspard}},
  \bibinfo{journal}{Phys. Rev. E} \textbf{\bibinfo{volume}{49}},
  \bibinfo{pages}{79} (\bibinfo{year}{1994}),
  \urlprefix\url{https://link.aps.org/doi/10.1103/PhysRevE.49.79}.

\bibitem[{\citenamefont{Jyoti}(2017)}]{Jy17}
\bibinfo{author}{\bibfnamefont{D.}~\bibnamefont{Jyoti}},
  \bibinfo{journal}{M.Sc. Thesis, Dartmouth College}  (\bibinfo{year}{2017}),
  \urlprefix\url{https://arxiv.org/abs/1711.01948}.

\bibitem[{\citenamefont{Rufeil-Fiori et~al.}(2009)\citenamefont{Rufeil-Fiori,
  S\'{a}nchez, Oliva, Pastawski, and Levstein}}]{RfSaPa09}
\bibinfo{author}{\bibfnamefont{E.}~\bibnamefont{Rufeil-Fiori}},
  \bibinfo{author}{\bibfnamefont{C.~M.} \bibnamefont{S\'{a}nchez}},
  \bibinfo{author}{\bibfnamefont{F.~Y.} \bibnamefont{Oliva}},
  \bibinfo{author}{\bibfnamefont{H.~M.} \bibnamefont{Pastawski}},
  \bibnamefont{and} \bibinfo{author}{\bibfnamefont{P.~R.}
  \bibnamefont{Levstein}}, \bibinfo{journal}{Phys. Rev. A}
  \textbf{\bibinfo{volume}{79}}, \bibinfo{pages}{032324}
  (\bibinfo{year}{2009}).

\bibitem[{\citenamefont{Orban and Bellemans}(1967)}]{BeOr67}
\bibinfo{author}{\bibfnamefont{J.}~\bibnamefont{Orban}} \bibnamefont{and}
  \bibinfo{author}{\bibfnamefont{A.}~\bibnamefont{Bellemans}},
  \bibinfo{journal}{Phys. Lett.} \textbf{\bibinfo{volume}{24A}},
  \bibinfo{pages}{620} (\bibinfo{year}{1967}).

\bibitem[{\citenamefont{Pinto et~al.}(2004)\citenamefont{Pinto, Medina, and
  Pastawski}}]{PiMPa04}
\bibinfo{author}{\bibfnamefont{R.}~\bibnamefont{Pinto}},
  \bibinfo{author}{\bibfnamefont{E.}~\bibnamefont{Medina}}, \bibnamefont{and}
  \bibinfo{author}{\bibfnamefont{H.~M.} \bibnamefont{Pastawski}},
  \bibinfo{journal}{BAPS March Meeting} \textbf{\bibinfo{volume}{2004}},
  \bibinfo{pages}{J22.001} (\bibinfo{year}{2004}),
  \urlprefix\url{http://flux.aps.org/meetings/YR04/MAR04/baps/abs/S3420001.html}.

\bibitem[{\citenamefont{Manfredi and Hervieux}(2008)}]{ManHe08}
\bibinfo{author}{\bibfnamefont{G.}~\bibnamefont{Manfredi}} \bibnamefont{and}
  \bibinfo{author}{\bibfnamefont{P.-A.} \bibnamefont{Hervieux}},
  \bibinfo{journal}{Phys. Rev. Lett.} \textbf{\bibinfo{volume}{100}},
  \bibinfo{pages}{050405} (\bibinfo{year}{2008}).

\bibitem[{\citenamefont{Fern\'andez-Alc\'azar and Pastawski}(2015)}]{FePa15}
\bibinfo{author}{\bibfnamefont{L.~J.} \bibnamefont{Fern\'andez-Alc\'azar}}
  \bibnamefont{and} \bibinfo{author}{\bibfnamefont{H.~M.}
  \bibnamefont{Pastawski}}, \bibinfo{journal}{Phys. Rev. A}
  \textbf{\bibinfo{volume}{91}}, \bibinfo{pages}{022117}
  (\bibinfo{year}{2015}).

\bibitem[{\citenamefont{Sánchez et~al.}(2007)\citenamefont{Sánchez,
  Pastawski, and Levstein}}]{SaPaLe07}
\bibinfo{author}{\bibfnamefont{C.~M.} \bibnamefont{Sánchez}},
  \bibinfo{author}{\bibfnamefont{H.~M.} \bibnamefont{Pastawski}},
  \bibnamefont{and} \bibinfo{author}{\bibfnamefont{P.~R.}
  \bibnamefont{Levstein}}, \bibinfo{journal}{Physica B: Condensed Matter}
  \textbf{\bibinfo{volume}{398B}}, \bibinfo{pages}{472 }
  (\bibinfo{year}{2007}).

\bibitem[{\citenamefont{Pastawski
  et~al.}(1995{\natexlab{b}})\citenamefont{Pastawski, Levstein, and
  Usaj}}]{PaLeU95}
\bibinfo{author}{\bibfnamefont{H.~M.} \bibnamefont{Pastawski}},
  \bibinfo{author}{\bibfnamefont{P.~R.} \bibnamefont{Levstein}},
  \bibnamefont{and} \bibinfo{author}{\bibfnamefont{G.}~\bibnamefont{Usaj}},
  \bibinfo{journal}{Phys. Rev. Lett.} \textbf{\bibinfo{volume}{75}},
  \bibinfo{pages}{4310} (\bibinfo{year}{1995}{\natexlab{b}}).

\bibitem[{\citenamefont{Pastawski et~al.}(1996)\citenamefont{Pastawski, Usaj,
  and Levstein}}]{PUL96}
\bibinfo{author}{\bibfnamefont{H.~M.} \bibnamefont{Pastawski}},
  \bibinfo{author}{\bibfnamefont{G.}~\bibnamefont{Usaj}}, \bibnamefont{and}
  \bibinfo{author}{\bibfnamefont{P.~R.} \bibnamefont{Levstein}},
  \bibinfo{journal}{Chem. Phys. Lett.} \textbf{\bibinfo{volume}{261}},
  \bibinfo{pages}{329} (\bibinfo{year}{1996}).

\bibitem[{\citenamefont{M\'adi et~al.}(1997)\citenamefont{M\'adi, Brutscher,
  Schulte-Herbr\"uggen, Br\"uschweiler, and Ernst}}]{Ma+Er97}
\bibinfo{author}{\bibfnamefont{Z.}~\bibnamefont{M\'adi}},
  \bibinfo{author}{\bibfnamefont{B.}~\bibnamefont{Brutscher}},
  \bibinfo{author}{\bibfnamefont{T.}~\bibnamefont{Schulte-Herbr\"uggen}},
  \bibinfo{author}{\bibfnamefont{R.}~\bibnamefont{Br\"uschweiler}},
  \bibnamefont{and} \bibinfo{author}{\bibfnamefont{R.}~\bibnamefont{Ernst}},
  \bibinfo{journal}{Chem. Phys. Lett.} \textbf{\bibinfo{volume}{268}},
  \bibinfo{pages}{300} (\bibinfo{year}{1997}).

\bibitem[{\citenamefont{Cappellaro et~al.}(2011)\citenamefont{Cappellaro,
  Viola, and Ramanathan}}]{CaViRa11}
\bibinfo{author}{\bibfnamefont{P.}~\bibnamefont{Cappellaro}},
  \bibinfo{author}{\bibfnamefont{L.}~\bibnamefont{Viola}}, \bibnamefont{and}
  \bibinfo{author}{\bibfnamefont{C.}~\bibnamefont{Ramanathan}},
  \bibinfo{journal}{Phys. Rev. A} \textbf{\bibinfo{volume}{83}},
  \bibinfo{pages}{032304} (\bibinfo{year}{2011}).

\bibitem[{\citenamefont{Pastawski et~al.}(2015)\citenamefont{Pastawski,
  Yoshida, Harlow, and Preskill}}]{HaPPY15}
\bibinfo{author}{\bibfnamefont{F.~M.} \bibnamefont{Pastawski}},
  \bibinfo{author}{\bibfnamefont{B.}~\bibnamefont{Yoshida}},
  \bibinfo{author}{\bibfnamefont{D.}~\bibnamefont{Harlow}}, \bibnamefont{and}
  \bibinfo{author}{\bibfnamefont{J.}~\bibnamefont{Preskill}},
  \bibinfo{journal}{J. High Energ. Phys.} \textbf{\bibinfo{volume}{149}},
  \bibinfo{pages}{2015} (\bibinfo{year}{2015}).

\bibitem[{\citenamefont{Horgan}(2020{\natexlab{b}})}]{Hor21}
\bibinfo{author}{\bibfnamefont{J.}~\bibnamefont{Horgan}},
  \bibinfo{journal}{Scientific American}  (\bibinfo{year}{2020}{\natexlab{b}}),
  \urlprefix\url{https://www.scientificamerican.com/article/will-quantum-computing-ever-live-up-to-its-hype/}.

\bibitem[{\citenamefont{Yan and Sinitsyn}(2020)}]{YaSi20}
\bibinfo{author}{\bibfnamefont{B.}~\bibnamefont{Yan}} \bibnamefont{and}
  \bibinfo{author}{\bibfnamefont{N.~A.} \bibnamefont{Sinitsyn}},
  \bibinfo{journal}{Phys. Rev. Lett.} \textbf{\bibinfo{volume}{125}},
  \bibinfo{pages}{040605} (\bibinfo{year}{2020}).

\bibitem[{\citenamefont{Sinitsyn and Yan}(2020)}]{YaSi20sa}
\bibinfo{author}{\bibfnamefont{N.~A.} \bibnamefont{Sinitsyn}} \bibnamefont{and}
  \bibinfo{author}{\bibfnamefont{B.}~\bibnamefont{Yan}},
  \bibinfo{journal}{Scientific American}  (\bibinfo{year}{2020}),
  \urlprefix\url{https://www.scientificamerican.com/article/the-quantum-butterfly-noneffect/}.

\bibitem[{\citenamefont{Trabesinger}(2012)}]{Tra12}
\bibinfo{author}{\bibfnamefont{A.}~\bibnamefont{Trabesinger}},
  \bibinfo{journal}{Nature Physics} \textbf{\bibinfo{volume}{8}},
  \bibinfo{pages}{263} (\bibinfo{year}{2012}).

\bibitem[{\citenamefont{Georgescu et~al.}(2014)\citenamefont{Georgescu, Ashhab,
  and Nori}}]{GANo14}
\bibinfo{author}{\bibfnamefont{I.~M.} \bibnamefont{Georgescu}},
  \bibinfo{author}{\bibfnamefont{S.}~\bibnamefont{Ashhab}}, \bibnamefont{and}
  \bibinfo{author}{\bibfnamefont{F.}~\bibnamefont{Nori}},
  \bibinfo{journal}{Rev. Mod. Phys.} \textbf{\bibinfo{volume}{86}},
  \bibinfo{pages}{153} (\bibinfo{year}{2014}).

\end{thebibliography}

\end{document}